\documentclass[11pt, a4paper]{article}
\usepackage{jcappub}

\usepackage{color}

\usepackage{booktabs}
\usepackage{multirow}
\usepackage{amssymb}
\usepackage[export]{adjustbox}
\usepackage{floatrow}
\newfloatcommand{capbtabbox}{table}[][3.5in]
\newfloatcommand{capbfigbox}{figure}[][2.3in]
\usepackage{blindtext}
\usepackage{amsmath}
\usepackage{booktabs}
\usepackage{aas_macros}
\usepackage[caption=false]{subfig}

\def\lsim{\mathrel{\raise.3ex\hbox{$<$\kern-.75em\lower1ex\hbox{$\sim$}}}}
\def\gsim{\mathrel{\raise.3ex\hbox{$>$\kern-.75em\lower1ex\hbox{$\sim$}}}}

\begin{document}

\title{A self-consistent model of cosmic-ray fluxes and positron excess: Roles of nearby pulsars and a sub-dominant source population}

\author{Jagdish C. Joshi,}
\emailAdd{jjagdish@uj.ac.za}
\author{Soebur Razzaque}
\emailAdd{srazzaque@uj.ac.za}

\affiliation{Department of Physics, University of Johannesburg,\\ P. O.
  Box 524, Auckland Park 2006, South Africa}

\abstract{The cosmic-ray positron flux calculated using the cosmic-ray nuclei interactions in our Galaxy cannot explain observed data above 10~GeV. An excess in the measured positron flux is therefore open to interpretation. Nearby pulsars, located within sub-kiloparsec range of the Solar system, are often invoked as plausible sources contributing to the excess. We show that an additional, sub-dominant population of sources together with the contributions from a few nearby pulsars can explain the latest positron excess data from the Alpha Magnetic Spectrometer (AMS). We simultaneously model, using the DRAGON code, propagation of cosmic-ray proton, Helium, electron and positron and fit their respective flux data.  Our fit to the Boron to Carbon ratio data gives a diffusion spectral index of 0.45, which is close to the Kraichnan turbulent spectrum.}

\maketitle

%%%%%%Section 1%%%%%%%
\section{Introduction}
%%%%%%%%%%%%%%%%%%%%%%

Our Galaxy is a continuous source of most of the cosmic-ray charged particles observed on Earth. These cosmic rays, which can be classified into primary and secondary particles as well as the matter particles (proton, Helium, electron, etc.) and antimatter particles (antiproton, positron, etc.), convey a wealth of information about their sources in the Galaxy and intervening medium.  Recent data from the balloon-borne detector {\it High Energy Antimatter Telescope} (HEAT)~\cite{Barwick:1995gv}, satellite-borne detectors {\it Payload for Antimatter Matter Exploration and Light-nuclei Astrophysics} (PAMELA)~\cite{Picozza:2006nm}, Fermi-{\it Large Area Telescope} (LAT)~\cite{Atwood:2009ez} as well as the {\it Alpha Magnetic Spectrometer} (AMS)~\cite{Aguilar:2013qda} onboard the International Space Station has led to significant development in understanding the properties of the cosmic ray sources and interstellar medium (ISM).

Observations of cosmic-ray positron flux by HEAT~\cite{Barwick:1997ig}, PAMELA~\cite{pamela_13_pos}, AMS~\cite{Aguilar:2013qda} and Fermi-LAT~\cite{fermi_pos_excess_12} have shown an excess in the $\approx 10$-500~GeV energy range, different than expected from propagation and interactions of cosmic-ray protons in the ISM.  This excess of positron fraction in the data has been used to understand its origin via dark matter particles~\cite{Cholis:2008hb,Bergstrom:2008gr,posexc_wdm,Cholis:2013psa} and via astrophysical sources \cite{hooper_ener_inj,occamrazor_profumo,Yuksel:2008rf,Venter:2015gga,pulsr_intr_zhang,posex_possintr,lin_yuanxj,Shaviv:2009bu,Kachelriess:2011qv,hawc_hooper17}. An astrophysical interpretation is  widely favored over dark matter annihilation models at present, although it requires additional mechanisms and/or sources of positrons. If positrons, which are produced inside the cosmic ray sources due to hadronic interactions, undergo acceleration inside these sources, then this process can explain the positron excess in the 10-100's of
GeV energy range~\cite{blasi_posi_excess}. In another shock wave model of nearby SNRs, the acceleration of secondary particles in the vicinity of the sources has been calculated for the explaination of AMS-02 results~\cite{subir_nsnr}. Nearby pulsars~\cite{hooper_ener_inj,occamrazor_profumo,Yuksel:2008rf,Venter:2015gga,pulsr_intr_zhang} and supernova remnants (SNRs)~\cite{Shaviv:2009bu,Kachelriess:2011qv} are well-motivated sources of positron flux based on gamma-ray observations. Most recently, very high-energy gamma-ray data from the {\it High Altitude Water Cherenkov} (HAWC) detector~\cite{Abeysekara:2017hyn} have been used to associate positron excess with nearby pulsars~\cite{hawc_hooper17}.    

A self-consistent model of cosmic-ray sources and propagation, however, needs to fit cosmic-ray nuclei and lepton, including positron excess, flux data simultaneously~\cite{gaggero2012cosmic}. Such self-consistent models also reduce degeneracy of
sources and their properties (e.g., pulsar and/or SNR age, spectral properties of primary cosmic rays, etc.) as well as ISM properties (e.g., magnetic field structure, diffusion coefficient and turbulence spectrum).
Recent data from the AMS-02~\cite{Aguilar:2013qda,bbyc_ams16} in the 10's of MeV to $\gtrsim 1$ TeV energy range poses new challenge to this picture. Moreover, PAMELA data collected earlier in the same energy
range~\cite{pamela_exper_one,pamela_exper_two,pamela_13_pos,bbycpamela_14}, specially for cosmic-ray leptons, are not always in agreement. In this paper we model cosmic-ray proton, Helium, Boron-to-Carbon ratio, electron,
positron and positron excess data self consistently using the publicly available DRAGON\footnote{We have used the 3D version of the DRAGON code available at \url{http://www.dragonproject.org/}}
code~\cite{Evoli:2016xgn} for propagation from cosmic-ray sources as well as using the analytic diffusion-loss equation~\cite{atoyan_ssc} for
very nearby pulsars. In addition to conventional cosmic-ray sources and nearby pulsars, we need a new population of cosmic electron and positron sources, distributed as the conventional sources but with much less power, for a successful interpretation of most fluxes observed on Earth.

The outline of the paper is as follows. We discuss the propagation of cosmic rays in the energy range of 100's of MeV to 10's of TeV using the DRAGON code and resulting proton and Helium fluxes as well as Boron to Carbon (B/C) ratio in Section 2. We calculate the electron-positron contribution of nearby sources and overall fits to electron, positron and positron excess data in Section 3. We summarize and discuss this work in Section 4.

%%%%%Section 2%%%%%%%%%%%%%%%%%%%%%%%%%%%%%%%%%%%%%%%%%%%%%%%
\section{Cosmic ray diffusion model and flux levels on Earth}
%%%%%%%%%%%%%%%%%%%%%%%%%%%%%%%%%%%%%%%%%%%%%%%%%%%%%%%%%%%%%
\label{sec:crdm}

Cosmic rays are injected into the interstellar medium (ISM) after diffusive acceleration at the shock regions of astrophysical objects~\cite{bell1_78,bell2_78,blandfr_chlr_87}, for example in supernova remnants.  Their propagation in the ISM can be revealed using the observational features of the secondary cosmic ray nuclei, gamma rays and radio emission from the Galaxy~\cite{rv_smp2007}.
Given the source distribution and injection spectrum of these particles, their propagation from the sources to the observers on Earth can be modelled by solving the transport equation~\cite{rv_smp2007,gaggero2012cosmic}.  Physical processes such as cosmic ray interactions in the Galactic medium, scattering of cosmic rays in the regular and turbulent Galactic magnetic fields, convection flow of cosmic rays in the Galactic winds and radioactive decay of the nuclei have been considered in the numerical solutions of transport equation.  One such numerical code, called DRAGON \cite{Evoli:2016xgn}, can solve the transport equation for cosmic ray propagation in the Galaxy in two geometries, namely a cylindrical geometry with azimuthal symmetry and a three-dimensional box.  The cosmic-ray nuclei interaction with the gas distribution in this work is based on the $\gamma$-ray observations of our Galaxy \cite{gas_model_galprop}.

%%%%%%%%%%%%%%%%%%%%%%%%
\subsection{Model setup}

In our model we considered a diffusive reacceleration with no convection scenario for propagation of cosmic rays using the three dimensional model in the DRAGON code.  We did-not consider convection process during the cosmic-ray transport in our Galaxy, as this process lowers the B/C ratio in the 1-10 GeV energy range which needs to be recovered by using large values of Alfven speed.  The magnetic field structure has been selected in the DRAGON code of type Pshirkov, in which the disk component has a value $2\,\mu$G and the halo component can be in a range 2-5 $\mu$G based on the radio synchrotron emission of the Galaxy~\cite{pshirkov11}. We have used the Galactic magnetic field disk component value as $2\,\mu$G, halo component value as $4\,\mu$G and the turbulent component value as $10\,\mu$G, respectively. 

The diffusion coefficient varies with particle rigidity $\rho$ and the Galaxy vertical height $z$, with a reference height $z_t$, as
\begin{equation}
D(\rho,z) = \beta^{\eta} D_0 \left( \frac{\rho}{\rho_0} \right)^{\delta} 
\exp\left( \frac{z}{z_t} \right)\,,
\label{diffcoeff}
\end{equation}
where the typical value of $D_0$ is $\sim 10^{28}$~cm$^2$/s.  The power $\eta$ on the particle speed $\beta$ accounts for the low energy uncertainties due to the particle propagation in the ISM~\cite{bernardo_aniso}. In this propagation scenario we calculate the secondary nuclei, namely the Boron to Carbon ratio~\cite{bbycpamela_14,bbyc_ams16} to fix the diffusion coefficient and the size of the Galactic halo~\cite{bbycratiosupp}.  To fit the B/C ratio one can tune the index $\delta$ in Eq.~(\ref{diffcoeff}) in a range 0.3-0.6~\cite{index_ref}.  In a recent AMS result it has been shown that above rigidity $\rho = 65$ GV, $\delta$ takes a value $0.333\pm 0.014~({\rm stat}) \pm 0.005~({\rm syst})$~\cite{bbyc_ams16}.  The rigidity break $\rho_0$ for the diffusion coefficient in Eq.~(\ref{diffcoeff}) is very useful to tune the B/C ratio in the 100 MeV-10 GeV range~\cite{gaggero2012cosmic}.  Below 10 GeV, the effect of solar modulation on the charged particles can be adopted to fit the observational data~\cite{glee_axf_solmod}.  The solar modulation can be modelled with a potential $\phi$, which can modify the interstellar cosmic ray spectrum by a factor
\begin{equation}
\epsilon(E_k,Z,A,m_Z) = \frac{(E_k + m_Z)^2 - m_Z^2} 
{\left( E_k + m_Z + \frac{Z|q|}{A}\phi \right)^2 - m_Z^2}\,,
\end{equation}
where $E_k$ is the particle kinetic energy, $Z$ is the atomic number, $A$ is the mass number and $m_Z$ is the nuclear mass.

%%%%%%%%%%%%%%%%%%%%%%%%%%%%%%%%%%%%%%%%%%%%%%%%%%
\subsection{Primary cosmic-ray source populations}

We assumed the cosmic-ray sources follow a Lorimer-type spatial distribution, which is based on the pulsar distribution in our Galaxy~\cite{lorimer06}.  This distribution with galactocentric radius $r$ and vertical height $z$ follows as,
\begin{equation}
g(r,z)=\left( \frac{r}{r_{\odot}} \right)^{1.9}
\exp\left(5\frac{r_{\odot} - r}{r_{\odot}}-\frac{|z|}{0.2} \right)\,,
\end{equation}
where $r_{\odot}=8.3$ kpc is the position of the solar system with respect to the Galactic center.  We have injected three primary particle populations, which follow the Lorimer distribution, into the ISM and propagate those using the DRAGON code.  The spectra of these particles in units of GeV$^{-1}$ m$^{-2}$ s$^{-1}$ sr$^{-1}$ are listed below.  
\begin{enumerate}
\item  {\em Cosmic-ray nuclei spectrum.} We assumed that the injection spectra of heavy nuclei and proton are the same.  We model those in the 0.1~GeV-10~TeV range, with a single break at 7~GV as,
\begin{align}
%\left\{
\begin{array}{ll}
n(\rho_A) = n_A \times \begin{cases}
\left(\frac{\rho_A}{7~{\rm GV}}\right)^{-2}; & \rho_A \leq 7~{\rm GV} \\
\left(\frac{\rho_A}{7~{\rm GV}}\right)^{-2.32}; & \rho_A > 7~{\rm GV}\,.
\end{cases}
\end{array}
%\right\} 
\label{prot_spec}
\end{align}
\item {\em Cosmic-ray electron spectrum.}  For electrons accelerated in the sources above 0.1~GeV we have assumed two breaks, at 6.2~GV and 85~GV, in the spectrum as,
\begin{align}
%\left\{
\begin{array}{ll}
n(\rho_{e^-}) = n_{e^-} \times \begin{cases}
\left( \frac{\rho_{e^-}}{6.2~{\rm GV}} \right)^{-1.91}; & \rho_{e^-} \leq 6.2~{\rm GV} \\
\left( \frac{\rho_{e^-}}{6.2~{\rm GV}} \right)^{-2.71}; & 6.2~{\rm GV} <\rho_{e^-} < 85~{\rm GV} \\
\left( \frac{85}{6.2} \right)^{-2.71} 
\left( \frac{\rho_{e^-}}{85~{\rm GV}} \right)^{-2.36}; & \rho_{e^-} \geq 85~{\rm GV}\,,
\end{cases}
\end{array}
%\right\} 
\label{elec_spec}
\end{align}
with an exponential folding energy of 10~TeV.
\item {\em Additional $e^{\pm}$ spectra.} We have also injected two additional populations of $e^-$ and $e^+$ from the sources, above 0.1~GeV, with the following spectrum,
\begin{align}
%\left\{
\begin{array}{ll}
n(\rho_{e^\pm}) = n_{e^\pm} \times \begin{cases}
\left( \frac{\rho_{e^\pm}}{3.8} \right)^{-1.85}; & \rho_{e^\pm} \leq 3.8~{\rm GV}\\
\left( \frac{\rho_{e^\pm}}{3.8} \right)^{-2.32}; & \rho_{e^\pm} > 3.8~{\rm GV}\,,
\end{cases}
\end{array}
%\right\} 
\label{add_spec}
\end{align}
with an exponential folding energy of 10~TeV.
\end{enumerate}
These particles are propagated in a Galactic halo of size $(x, y, z) = (12~{\rm kpc}, 12~{\rm kpc}, 8~{\rm kpc})$.  We use the diffusion coefficient in Eq.~(\ref{diffcoeff}) with values $D_0=2.8 \times 10^{28}$~cm$^2$/s, $\rho_0 = 4.5$~GV and $\delta=0.45$. We have also used Alfven speed $v_A=19$~km/s. 
The electron spectrum has two breaks, which has been discussed in earlier works, for the interstellar cosmic-ray spectrum, where the injected spectral index is harder after the second break \cite{twobreaks}.

%%%%%%%%%%%%%%%%%%%%%%%%%%%%%%%%%%%%%%%%%%%%%%%%%%%
\subsection{Fluxes of nuclei and cosmic-ray powers}

\begin{figure}[htbp]
\centering
%   \subfloat[label 1]{{\includegraphics[width=7cm]{BC3D.eps}}}%
\includegraphics[width=7cm]{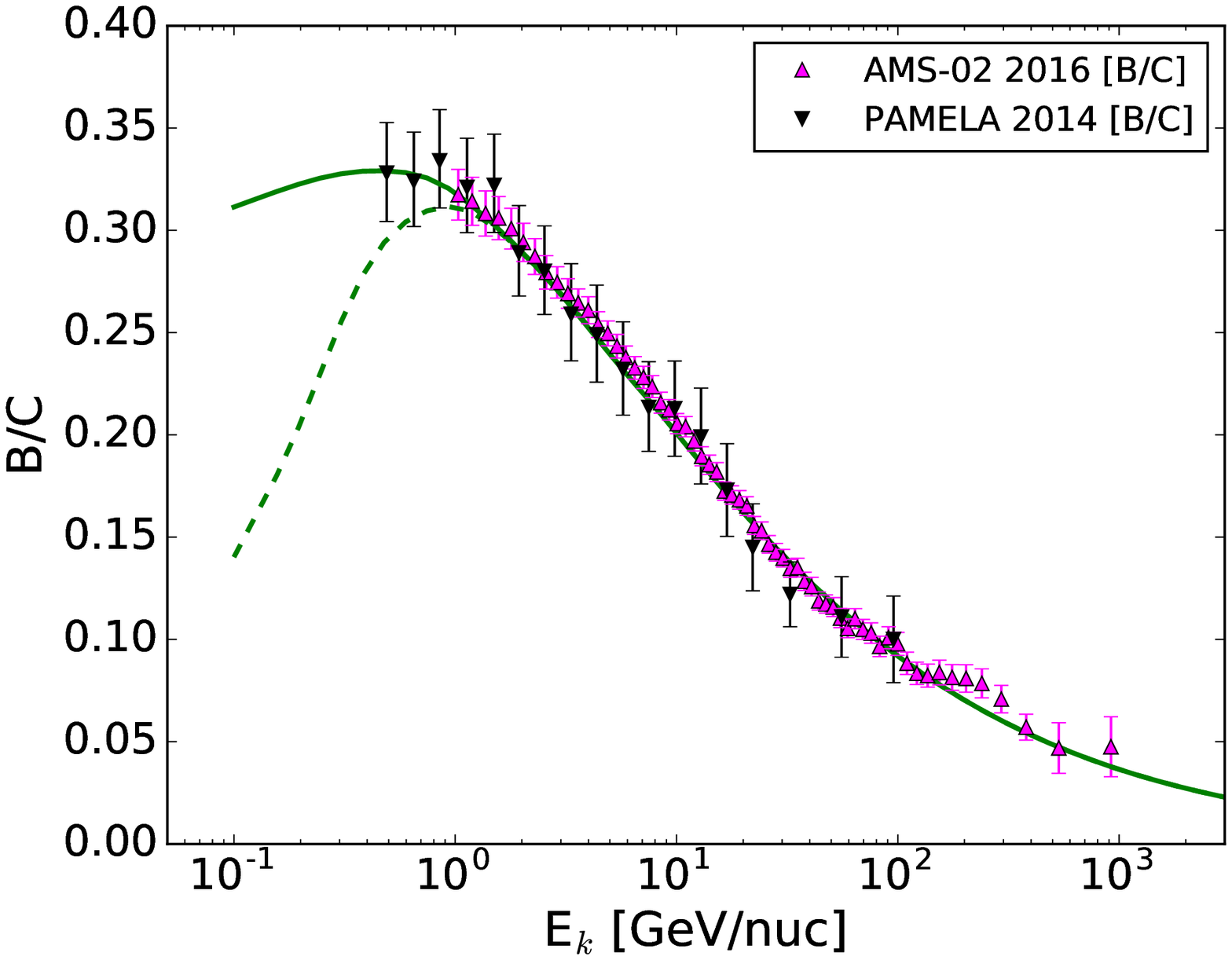}
\qquad
% \subfloat[label 2]{{\includegraphics[width=7cm]{cosray.eps}}}%
\includegraphics[width=7cm]{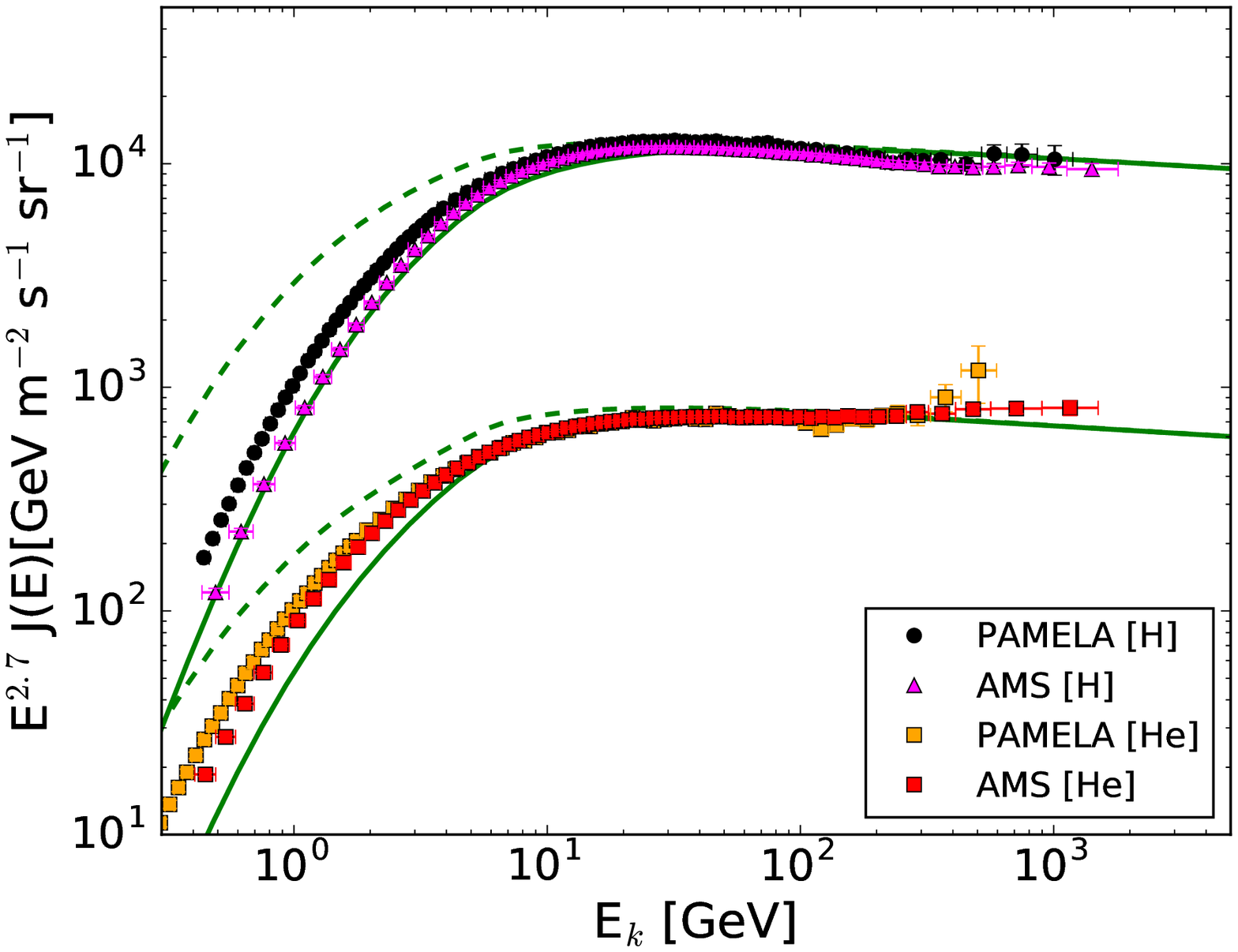}
\caption{{\em Left panel:} B/C ratio calculation using the DRAGON code and plotted against the AMS~\cite{bbyc_ams16} and PAMELA~\cite{bbycpamela_14} data.  {\em Right panel:} Proton and Helium fluxes plotted against the PAMELA~\cite{pamela_proton} and AMS data~\cite{ams_proton,heams2}.  The solid (dashed) lines are fluxes with (without) solar modulation taken into account.}
\label{fig:nuclei}%
\end{figure}

We have plotted the Boron to Carbon flux ratios in the left panel of Fig.~\ref{fig:nuclei}, while in the right panel we have plotted the proton and Helium fluxes along with observed data.\footnote{All data are taken from the database~\cite{database} unless otherwise specified.}  The solid (dotted) lines are fluxes after (before) taking into account solar modulation.  Our choice of model parameters fits the B/C ratio data from both the AMS-02~\cite{bbyc_ams16} and PAMELA~\cite{bbycpamela_14} quite well in the whole observed energy range. This requires the index of the diffusion coefficient to be $\delta = 0.45$, close to the Kraichnan turbulence spectrum rather than the Kolmogorov turbulence spectrum of $\approx 0.33$ above 65 GV, reported in~\cite{bbyc_ams16}.  

The proton flux fits the AMS data very well but shows a departure from the PAMELA data at low energies.  We think this discrepancy comes due to a difference in the solar activity in the epochs of these two observations.  The Helium flux fits the AMS and PAMELA data quite well at high energies but deviates from both, below $\approx 4$~GeV.  Again, solar modulation effect is a plausible explanation for this.

We calculate the total powers in cosmic-ray components, using the spectra in Eqs.~(\ref{prot_spec}); (\ref{elec_spec}) and (\ref{add_spec}), by injecting and propagating these particles using the DRAGON code for $T = 64$ Million years in a volume $V=1.15\times10^3 \rm~kpc^3$ of the Galaxy, as
\begin{equation}
P = \frac{V}{T} \frac{4\pi}{c} \int_{E_1}^{E_2} n(E) E dE \,.
\end{equation}
Here $E_1$ and $E_2$ are the two limiting energies in the spectrum.  In our calculations, the approximate total power required in the protons is $10^{56}$~GeV/Myr, in primary electrons is $2.8 \times 10^{54}$~GeV/Myr and in the additional $e^\pm$ populations is $1.5 \times 10^{53}$~GeV/Myr.

%%%%%%%%%%%%%%%%%%%%Section 3%%%%%%%%%%%%%%%%%%%%%%
\section{Contributions from nearby sub-kpc pulsars}
%%%%%%%%%%%%%%%%%%%%%%%%%%%%%%%%%%%%%%%%%%%%%%%%%%%

Acceleration of charged particles in pulsars can be explained using the polar cap and outer gap models~\cite{ruderman_86_pcap,baring_pcpa}. If the pulsar age is in the range of 40-50 kyr, then it remains with the parent supernova remnant \cite{blasi_40_50kyr}, which means in the younger pulsars (below $\sim 50$ kyr) we can see emission of electrons due to their SNRs, and above $\sim 50$ kyr the emission of electron-positron pairs would occur from its pulsar wind nebula (PWN). The time scale for the $e^--e^+$ pair escape from the PWN is approximately $10^5$ years, when the ISM pressure will disturb the PWN structure and an enhanced flux of accelerated particles will be injected into the ISM~\cite{venter_08}. 
We have selected nearby pulsars from the Australia Telescope National Facility (ATNF) catalogue\footnote{\url{http://www.atnf.csiro.au/people/pulsar/psrcat/}} with parameters taken from there as well as from~\cite{displs_min_1,b105552dis_2} and we have listed those in Table~\ref{tab:pulcat}.

The distance to a pulsar can have model dependent outcomes, for example in case of B1055-52, based on Galactic free electron density model the distance can be 1.53 kpc \cite{taylr_cordes_93} or $0.73 \pm 0.15$ kpc as estimated in \cite{cordes_lazio}. We have taken the distance to be $0.35 \pm 0.15$ kpc as estimated in a recent work \cite{b105552dis_2} using observational features in optical-UV and X-ray thermal components of B1055-52 pulsar. The energy injected by a pulsar in time $t$, can be estimated using the formalism provided in \cite{hooper_ener_inj}. For a pulsar with initial period $P_0$ (in seconds), radius $R_s$ (in km), and surface magnetic field
$B_s$ (in Gauss), age $t$ (in years), the total rotational energy $E_{\rm tot}(t)$ in ergs is given by

\begin{equation}
E_{\rm tot}(t) = 6 \times 10^{43}   \eta_{e^\pm}  P_0^{-4} \left (\frac{R_s}{10}\right)^6 \left(\frac{B_s}{10^{12}}\right)^2 \left(\frac{t}{10^{5}}\right) \frac{1}{(1+\frac{t}{\tau_0})}
\label{pulsarE}
\end{equation}
\noindent
Here $\tau_0 \sim 10^4$ is the luminosity decay time for mature pulsars. If $\eta_{e^\pm}$ is the fraction of the total energy that goes into $e^--e^+$ pair emissions then the contribution of nearby pulsars for the pair emission
can be estimated. We have taken $\eta_{e^\pm} = 1.0$ in Table~\ref{tab:pulcat}. In Equation \ref{pulsarE}, the total energy is very sensitive parameter of the radius of the pulsar, and we take a value of 10 km in the estimation
for the rotational energy. The initial period $P_0$, for these pulsars has been estimated using $P_0=P(\tau_0/t)^{1/2}$, here $P$ is the period of a pulsar at an age $t$ \cite{hooper_ener_inj}.

\begin{table}[htbp]
\centering
\caption{Pulsar parameters based on the ATNF catalogue. The contribution of these pulsars to the electron-positron fluxes in the 100 GeV-1 TeV has been calculated in this work.}
\vskip 0.5cm
\vskip 0.25cm
\begin{tabular}{|l|l|l|l|l|l|}
\hline
Name & Dist (kpc) & Age (yr) & Mag-field (G)& P (ms) & $E_{\rm tot}(t)$ (erg)\\
\hline
% B0450-18 & 0.40 & 1.51 $\times 10^6$ & 1.4 $\times 10^{33}$ \\
% J0611+1436 & 0.89 & 1.07 $\times 10^6$ & 8.0 $\times 10^{33}$ \\
J0659+1414 (Monogem) & $0.29^{+0.03}_{-0.03}$ & 1.1 $\times 10^5$&4.66 $\times 10^{12}$ &384& $6.6 \times 10^{47}$\\
J0633+1746 (Geminga) & $0.25^{+0.23}_{-0.08}$ & 3.42 $\times 10^5$& 1.63$\times 10^{12}$& 237&$5.7 \times 10^{48}$ \\
%J0745-5353 & 0.57 & 1.25 $\times 10^6$ &0.78 $\times 10^{12}$ & $2.9\times 10^{50}$\\
%J0835-3707 & 0.55 & 8.77 $\times 10^5$ & 2.4 $\times 10^{33}$ \\
% B0905-51 & 0.34 & 2.19 $\times 10^6$ & 4.4 $\times 10^{33}$ \\
% B0922-52 & 0.51 & 3.33 $\times 10^5$ & 3.4 $\times 10^{33}$ \\
% B0940-55 & 0.30 & 4.61 $\times 10^5$ & 3.1 $\times 10^{33}$ \\
% B0941-56 & 0.41 & 3.23 $\times 10^5$ & 3.0 $\times 10^{33}$ \\
% J0945-4833 & 0.35 & 1.09 $\times 10^6$ & 5.2 $\times 10^{33}$ \\
%J0954-5430 & 0.43 & 1.71 $\times 10^5$ &4.61 $\times 10^{12}$ &472&$7.1\times 10^{47}$ \\
% J0957-5432 & 0.45 & 1.66 $\times 10^6$ & 9.1 $\times 10^{33}$ \\
% J1000-5149 & 0.13 & 4.22 $\times 10^6$ & 2.3 $\times 10^{33}$ \\
%J1003-4747 & 0.37 & 2.2 $\times 10^5$ &2.63 $\times 10^{12}$ &307&$1.3\times 10^{49}$\\
J1057-5226 (B1055-52) & $0.35^{+0.15}_{-0.15}$ &5.35 $\times 10^5$ &1.09 $\times 10^{12}$ &197&$1.3 \times 10^{49}$\\
%J1741-2054 & 0.3 & 3.86 $\times 10^5$ &2.68 $\times 10^{12}$&414 & $2.1\times 10^{48}$\\
% B0833-45 (Vela) & 0.28 &$\bm{ 1.13 \times 10^4}$ & $\bm{1.6 \times 10^{36}}$& \\
% J0940-5428 & 0.38 & $\bm{4.22 \times 10^4}$ & $\bm {1.9 \times 10^{36}}$& \\
% B1737-30 & 0.4 &$\bm{ 2.06 \times 10^4}$ & $\bm{8.2 \times 10^{34}}$ \\
% J1741-2054 & 0.3 & 3.86 $\times 10^5$ & 9.5 $\times 10^{33}$ \\
% B1742-30 & 0.2 & 5.46 $\times 10^5$ & 8.5 $\times 10^{33}$ \\
% B1749-28 & 0.2 & 1.1 $\times 10^6$ & 1.8 $\times 10^{33}$ \\
% J1755-0903 & 0.23 & 3.87 $\times 10^6$ & 4.4 $\times 10^{33}$ \\
% B1818-04 & 0.30 & 1.5 $\times 10^6$ & 1.2 $\times 10^{33}$ \\
% J1908+0734 & 0.67 & 4.08 $\times 10^6$ & 3.4 $\times 10^{33}$ \\
% J1918+1541 & 0.38 & 2.31 $\times 10^6$ & 2.0 $\times 10^{33}$ \\
% B1929+10 & 0.31 & 3.1 $\times 10^6$ & 3.9 $\times 10^{33}$ \\
% B2334+61 (SNR:G114.3+0.3) & 0.70 &$\bm{ 4.06 \times 10^4}$ & $\bm{6.3 \times 10^{34}}$ \\
\hline
\end{tabular}
\label{tab:pulcat}
\end{table}

%%%%%%%%%%%%%%%%%%%%%%%%%%%%%%%%%%%%%%%%%%%%%%%%%%%%%%%%%
\subsection{Electron-positron fluxes from nearby sources}

The nearby source contribution to the cosmic-ray nuclei fluxes is negligible compared to the diffuse background from the Galaxy~\cite{lingenfelter_69}, but for electrons and positrons there can be sizeable contributions to the observed fluxes from nearby sources.

In the energy range above $\sim 1$~GeV, relativistic electrons (or positrons) loose their energy mainly via synchrotron and inverse Compton interactions~\cite{blu_gould1970,kobayashi_10gev}. If these charged particles are emitted from sub-kpc or nearby cosmic-ray sources, then they can propagate through the interstellar radiation field (ISRF) and Galactic magnetic field, loose their energy and still contribute significantly to the fluxes on Earth. 

The energy losses in the ISRF, with energy density $U_{\rm ISRF}$, and magnetic field, with energy density $U_B$, can be written as 
\begin{equation}
-\frac{dE_e}{dt} = \frac{4}{3}\sigma_Tc\gamma_e^2 U_{\rm T}=b_0 E_e^2\,,
\end{equation}
where $b_0 = 1.06\times10^{-16} U_{\rm T}/(1~{\rm eV}/{\rm cm}^3)$~GeV$^{-1}$~s$^{-1}$ and $U_{\rm T} = U_{\rm ISRF}+U_B$.  We have taken $U_{\rm T} = 1.32~{\rm eV}/{\rm cm}^3$ to match with the energy loss rate coefficient $b_0 = 1.4\times10^{-16}$  GeV$^{-1}$s$^{-1}$ used in the DRAGON code~\cite{Evoli:2016xgn}. If an electron with initial energy $E_0$ travels through these radiation and magnetic fields then its energy at any time $t$ is given by
\begin{equation}
E_e(t)= \frac{E_0}{1+b_0E_0t}\,.
\label{Eet}
\end{equation}
An electron with observed energy $E_e$ on Earth must be emitted from its source at a time $t \sim 1/b_0E_e = 2.0 \times 10^5 (E_e/{\rm TeV})^{-1}$ years, from Eq.~(\ref{Eet}) with $E_0\gg E_e$.  Within this time scale the distance travelled by an electron of energy in the 100's of GeV to TeV range is approximately $\sqrt{2 D(E)t}$, which is in the range of sub-kpc \cite{berezinskii_buk}.  Here $D(E)$ is a diffusion coefficient that depends on energy only as we will discuss shortly.  As a result, the study of nearby cosmic-ray electron-positron sources has always been considered for the estimation of GeV-TeV electron-positron fluxes on Earth \cite{shen_pulsar70,atoyan_ssc,venter_08,occamrazor_profumo,malyshev_pulsarageetc,hawc_hooper17}. 

Particle injection in the form of a burst into the ISM from an individual source located at a distance $r$ with age $t$ can be described as,
\begin{equation}
Q(E,t,r) = Q_0 \left( \frac{E}{\rm GeV} \right)^{-\Gamma} \exp\left(-\frac{E}{E_c} \right) 
\delta(t-t_0) \delta(r) \,,
\label{epluminusspectra}
\end{equation}
where $Q(E,t,r)$ is the cosmic-ray electron (or positron) emissivity, i.e., density per unit energy~\cite{atoyan_ssc}.  Also, $Q_0$ and $\Gamma$ are normalization and index for the power-law part of the spectrum and $E_{\rm c}$ is an exponential folding energy.  A comparison of the amount of spin-down energy, in Eq.~(\ref{pulsarE}), going into $e^\pm$ pair emissions with the integrated particle injection spectrum, in Eq.~(\ref{epluminusspectra}), provides the numerical estimation of $Q_0$.  In Eq.~(\ref{epluminusspectra}) $t_0$ is a time delay for emission and we have considered it 85~kyr for the optimized electron-positron emission from the pulsar-wind nebula (PWN) of a pulsar.
The choice of the injection time is based on the fact that PWN emission starts dominating within 100 kyr of the pulsar age \cite{venter_08}.

With the above source spectrum and considering only energy-dependent diffusion in the ISM and energy losses due to the ISRF and magnetic field one can solve the diffusion-loss equation to
calculate the particle density per unit energy on Earth as~\cite{atoyan_ssc,grasso_delta},
\begin{equation}
%\begin{split}
N_e(E,t,r) = \frac{Q_0}{\pi^{3/2}r_{\rm diff}^3} \left[1-\frac{E}{E_{\rm max}(t)}\right]^{\Gamma-2} 
\exp\left[-\left(\frac{r}{r_{\rm diff}}\right)^2 \right]
\exp\left[ -\frac{E/E_c}{1-E/E_{\rm max} (t)} \right]\,.
%\end{split}
\label{single_spec}
\end{equation}
The function $N_e$ is equal to zero above a value $E_{\rm max} =1/b_0 (t-t_0)$ and can be used to calculate the differential flux $J=cN_e/4\pi$ from the nearby sources.  The diffusion length scale in Eq.~(\ref{single_spec}) is defined as,
\begin{equation}
r_{\rm diff}(E,t) = 2 \sqrt{D(E)t \frac{1-[1-E/E_{\rm max}(t)]^{1-\delta}}
{(1-\delta)E/E_{\rm max}(t)}}\,,
\label{diff_length}
\end{equation}
where the diffusion coefficient $D(E)$ depends only on the particle energy.

For pulsars, the flux on Earth in the energy above 100~GeV depends on the age and distance to these sources~\cite{malyshev_pulsarageetc}.  Using Eqs.~(\ref{single_spec}) and (\ref{diff_length}) one can show that the observed flux on Earth depends on the pulsar age $t$ and its distance $r$ as $N_e(E,t,r) \propto t^{-3/2}\exp(-r^2/t^2)$.  Next we discuss the individual pulsars we have considered for electron-positron fluxes.

%%%%%%%%%%%%%%%%%%%%%%%%%%%%%%%%%%%%%%%%%%%%%%%%%%%%%%%%%%%%%%
\subsection{Electron-positron flux levels and positron excess}

\begin{figure}[htbp]
\centering
%\subfloat[label 1]{{\includegraphics[width=7cm]{leptons.eps} }}%
\includegraphics[width=7.cm]{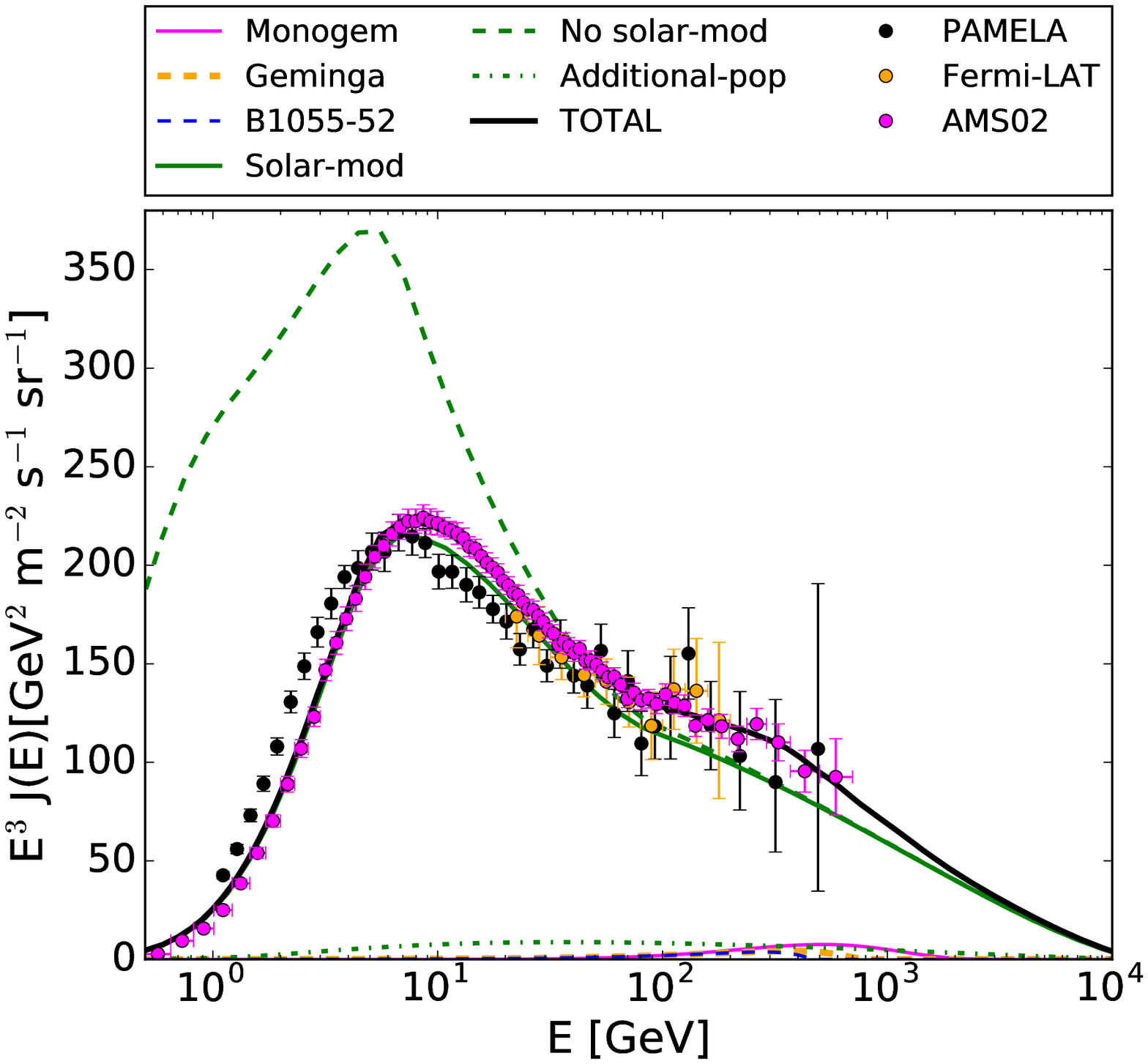}
\qquad
%\subfloat[label 2]{{\includegraphics[width=7cm]{lepton_pos.eps} }}%
\includegraphics[width=7cm]{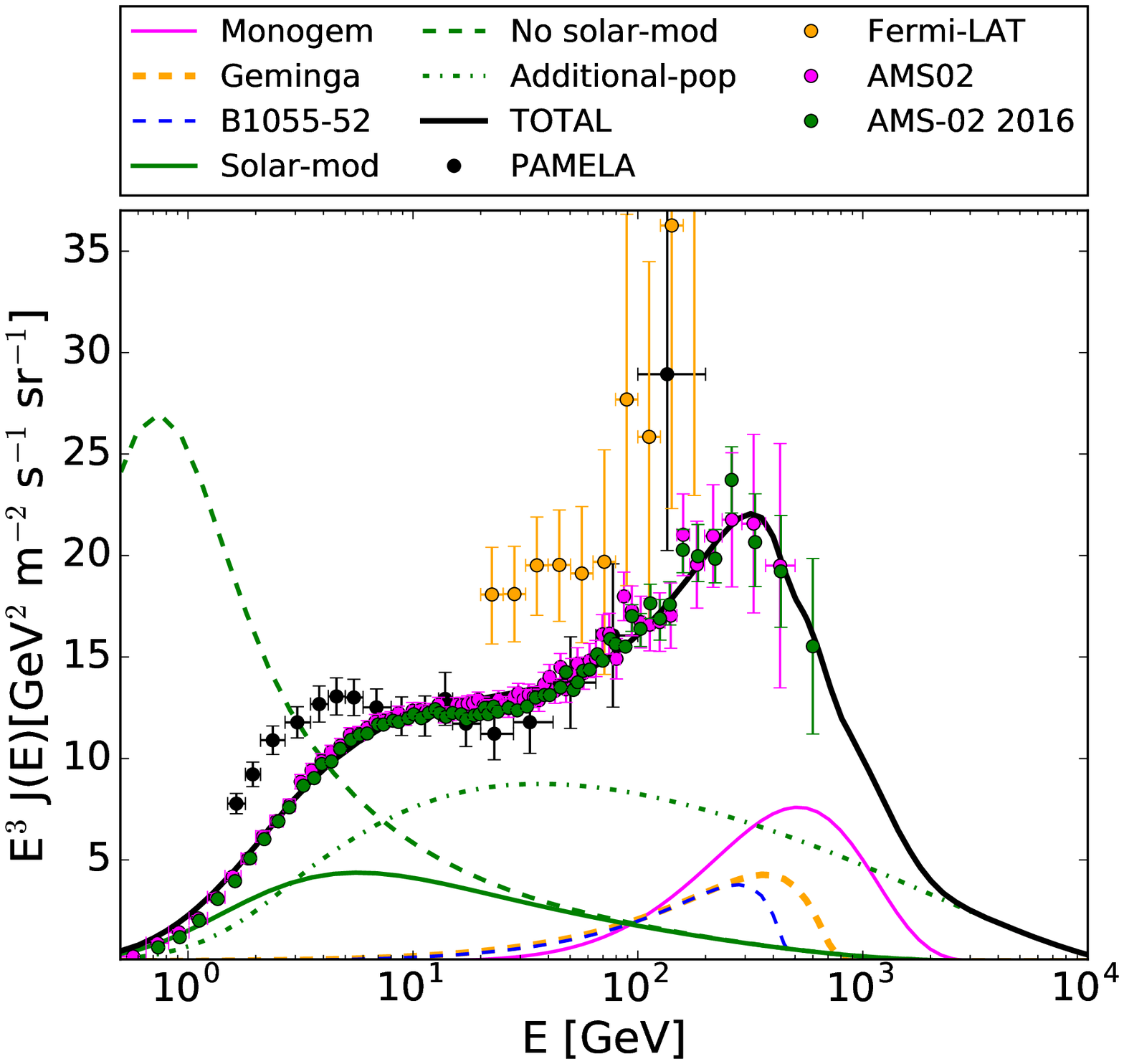}
\caption{{\em Left panel:} Electron flux data from PAMELA~\cite{pamela_exper_two}, Fermi-LAT~\cite{fermi_pos_excess_12} and AMS-02~\cite{ams02_elec}.  {\em Right panel:} Positron Flux data from PAMELA~\cite{pamela_13_pos}, Fermi-LAT~\cite{fermi_pos_excess_12} and AMS-02~\cite{ams02_elec}, both for an injection time 85 kyr. Also shown are our model fluxes with black solid line for the total flux in both panels.  See main text for more details.}
\label{fig:leptons}
\end{figure}

We have plotted in Fig.~\ref{fig:leptons} our model fit to the cosmic-ray electron (left panel) and positron (right panel) flux data.  The contribution from conventional primary electron sources, following Lorimer distribution, in the Galaxy is shown as green solid (dashed) line after (before) taking into account solar modulation.  For positrons, the solid (dashed) green line corresponds to the contribution from primary cosmic-ray nuclei interactions in the Galaxy.  Note that this component is significant only at very low energies.  The contributions from an additional primary $e^\pm$ sources, which follow a Lorimer-type distribution, are shown in both panels with green dot-dashed lines, after taking into account solar modulation.  While for electrons this component is negligible, for positrons this is crucial to fit data.  We have shown contributions from the nearby pulsars, in both panels, whose contributions are crucial to fit high energy, $\gtrsim 100$~GeV, data.  The black solid lines in both panels correspond to the total model flux.  Note that we fit the AMS-02 data~\cite{ams02_elec} with our model very well over the whole energy range.

\begin{figure}[htbp]
\centering
%\subfloat[label 1]{{\includegraphics[width=7cm]{leptons.eps} }}%
\includegraphics[width=7.cm]{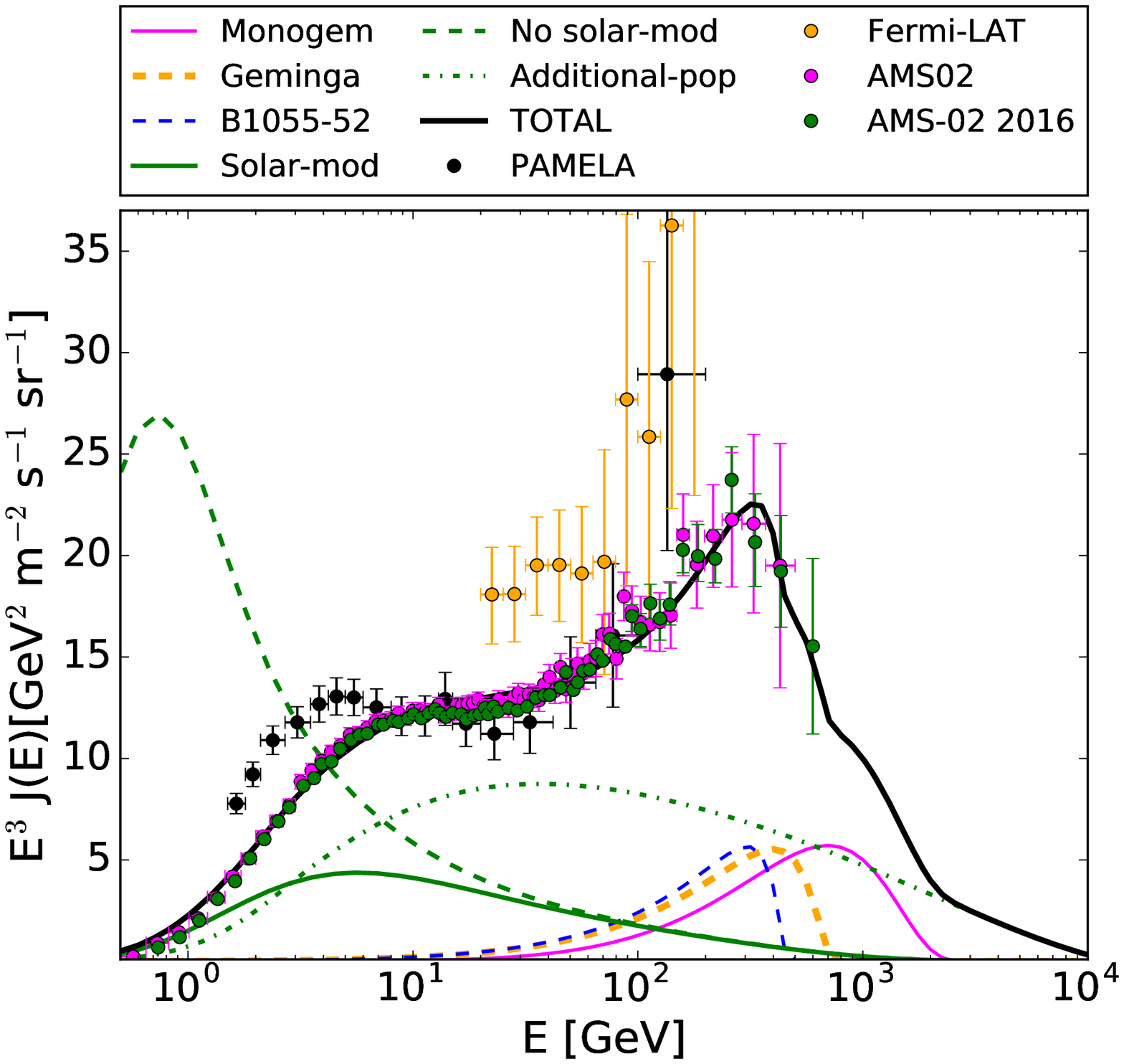}
\qquad
%\subfloat[label 2]{{\includegraphics[width=7cm]{lepton_pos.eps} }}%
\includegraphics[width=7cm]{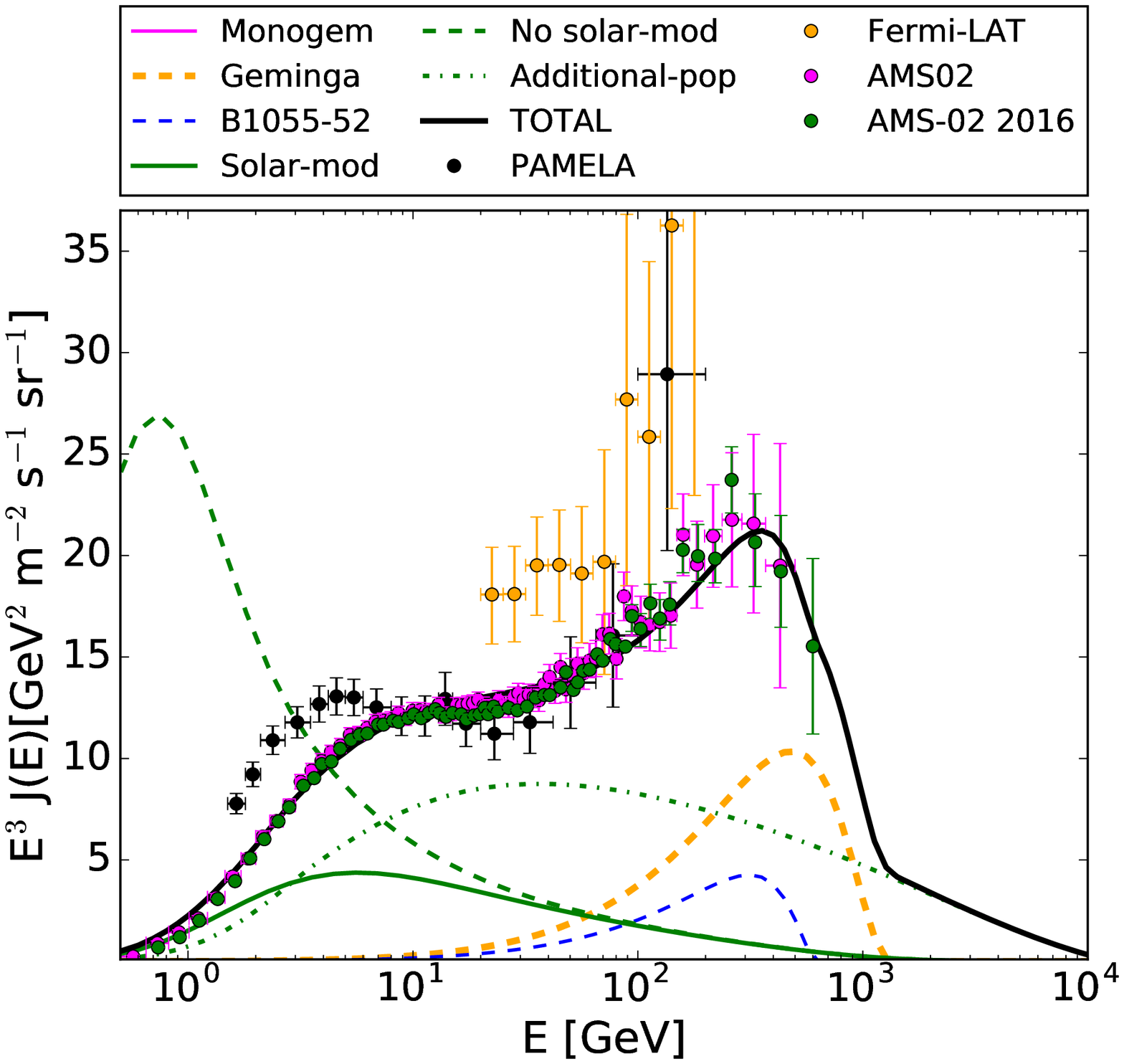}
\caption{{\em Left panel:} Positron Flux data from PAMELA~\cite{pamela_13_pos}, Fermi-LAT~\cite{fermi_pos_excess_12} and AMS-02~\cite{ams02_elec} for an injection time 50 kyr years.  {\em Right panel:} Positron Flux data from PAMELA~\cite{pamela_13_pos}, Fermi-LAT~\cite{fermi_pos_excess_12} and AMS-02~\cite{ams02_elec} for an injection time 200 kyr. Also shown are our model fluxes with black solid line for the total flux in both panels.}
\label{fig:inj_ref}
\end{figure}
\begin{table}[htbp]
\centering
\caption{Nearby pulsar injection parameters used in this calculation, here $\Gamma$ is the index of the power law spectrum, $E_c$ is the exponential folding energy and $t_0$ is the injection time, defined in Equation \ref{epluminusspectra}.
The corresponding contribution of these pulsars w.r.t. change in injection time $t_0$ is shown in Fig. \ref{fig:inj_ref} and in the right panel of Fig. \ref{fig:leptons}. For $t_0$=200 kyr, the positron flux from Monogem is indicated by (none), because its lifetime (100 kyr), is shorter than the injection time ($t_0$=200 kyr).}
\vskip 0.5cm
\vskip 0.25cm
\begin{tabular}{|l|l|l|l|}
\hline
~~~~Pulsar name &($\Gamma$,$\eta_e^{\pm}$, $E_c$(GeV))& ($\Gamma$,$\eta_e^{\pm}$,$E_c$(GeV)) & ($\Gamma$,$\eta_e^{\pm}$,$E_c$(GeV))\\
 & $t_0$=50 kyr & $t_0$=85 kyr & $t_0$=200 kyr\\

\hline
% B0450-18 & 0.40 & 1.51 $\times 10^6$ & 1.4 $\times 10^{33}$ \\
% J0611+1436 & 0.89 & 1.07 $\times 10^6$ & 8.0 $\times 10^{33}$ \\
J0659+1414 (Monogem) & (1.4, $7 \%$, 900) & (1.6, $8 \%$, 500) &~~~(none)\\
J0633+1746 (Geminga) &(1.4, $8 \%$, 1200) & (1.5, $6 \%$, 1000)&(1.4, $5 \%$, 850) \\
%J0745-5353 & 0.57 & 1.25 $\times 10^6$ &0.78 $\times 10^{12}$ & $2.9\times 10^{50}$\\
%J0835-3707 & 0.55 & 8.77 $\times 10^5$ & 2.4 $\times 10^{33}$ \\
% B0905-51 & 0.34 & 2.19 $\times 10^6$ & 4.4 $\times 10^{33}$ \\
% B0922-52 & 0.51 & 3.33 $\times 10^5$ & 3.4 $\times 10^{33}$ \\
% B0940-55 & 0.30 & 4.61 $\times 10^5$ & 3.1 $\times 10^{33}$ \\
% B0941-56 & 0.41 & 3.23 $\times 10^5$ & 3.0 $\times 10^{33}$ \\
% J0945-4833 & 0.35 & 1.09 $\times 10^6$ & 5.2 $\times 10^{33}$ \\
%J0954-5430 & (1.65,0.2) & (1.7,0.15) &(1.9,0.15) \\
% J0957-5432 & 0.45 & 1.66 $\times 10^6$ & 9.1 $\times 10^{33}$ \\
% J1000-5149 & 0.13 & 4.22 $\times 10^6$ & 2.3 $\times 10^{33}$ \\
%J1003-4747 &(1.65,0.2)& (1.7,0.15) &(1.9,0.15)\\
J1057-5226 (B1055-52) &(1.4, $9 \%$, 1400)&(1.45, $6 \%$, 1000)&(1.4, $4 \%$, 800) \\
%J1741-2054 & (1.65,0.2) & (1.7,0.15) &(1.9,0.15)\\
% B0833-45 (Vela) & 0.28 &$\bm{ 1.13 \times 10^4}$ & $\bm{1.6 \times 10^{36}}$& \\
% J0940-5428 & 0.38 & $\bm{4.22 \times 10^4}$ & $\bm {1.9 \times 10^{36}}$& \\
% B1737-30 & 0.4 &$\bm{ 2.06 \times 10^4}$ & $\bm{8.2 \times 10^{34}}$ \\
% J1741-2054 & 0.3 & 3.86 $\times 10^5$ & 9.5 $\times 10^{33}$ \\
% B1742-30 & 0.2 & 5.46 $\times 10^5$ & 8.5 $\times 10^{33}$ \\
% B1749-28 & 0.2 & 1.1 $\times 10^6$ & 1.8 $\times 10^{33}$ \\
% J1755-0903 & 0.23 & 3.87 $\times 10^6$ & 4.4 $\times 10^{33}$ \\
% B1818-04 & 0.30 & 1.5 $\times 10^6$ & 1.2 $\times 10^{33}$ \\
% J1908+0734 & 0.67 & 4.08 $\times 10^6$ & 3.4 $\times 10^{33}$ \\
% J1918+1541 & 0.38 & 2.31 $\times 10^6$ & 2.0 $\times 10^{33}$ \\
% B1929+10 & 0.31 & 3.1 $\times 10^6$ & 3.9 $\times 10^{33}$ \\
% B2334+61 (SNR:G114.3+0.3) & 0.70 &$\bm{ 4.06 \times 10^4}$ & $\bm{6.3 \times 10^{34}}$ \\
\hline
\end{tabular}
\label{tab:pul_feather}
\end{table}

\begin{figure}[htbp]
\centering
%\subfloat[label 1]{{\includegraphics[width=7cm]{leptons.eps} }}%
\includegraphics[width=7.2cm]{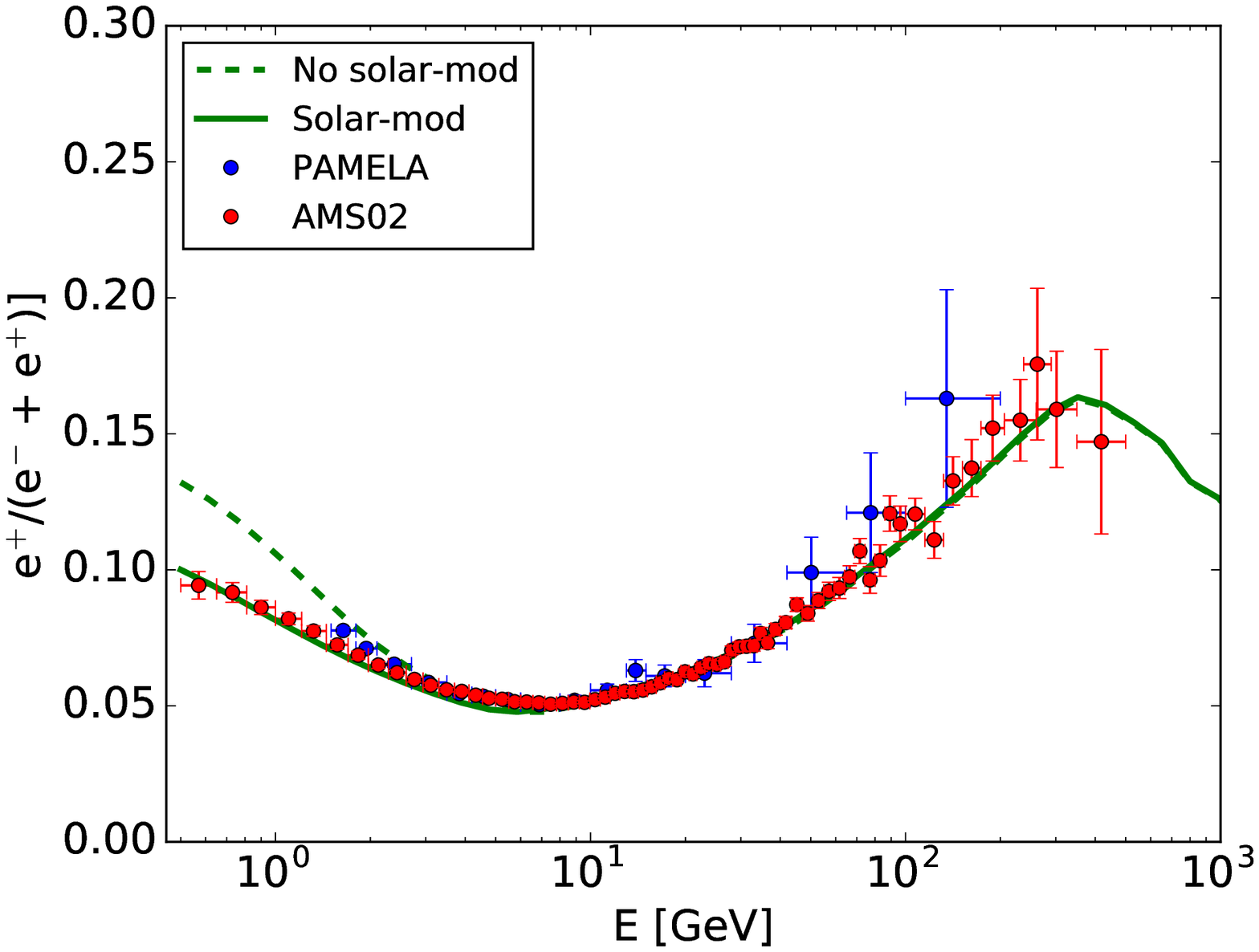}
\qquad
%\subfloat[label 2]{{\includegraphics[width=7cm]{lepton_pos.eps} }}%
\includegraphics[width=7cm]{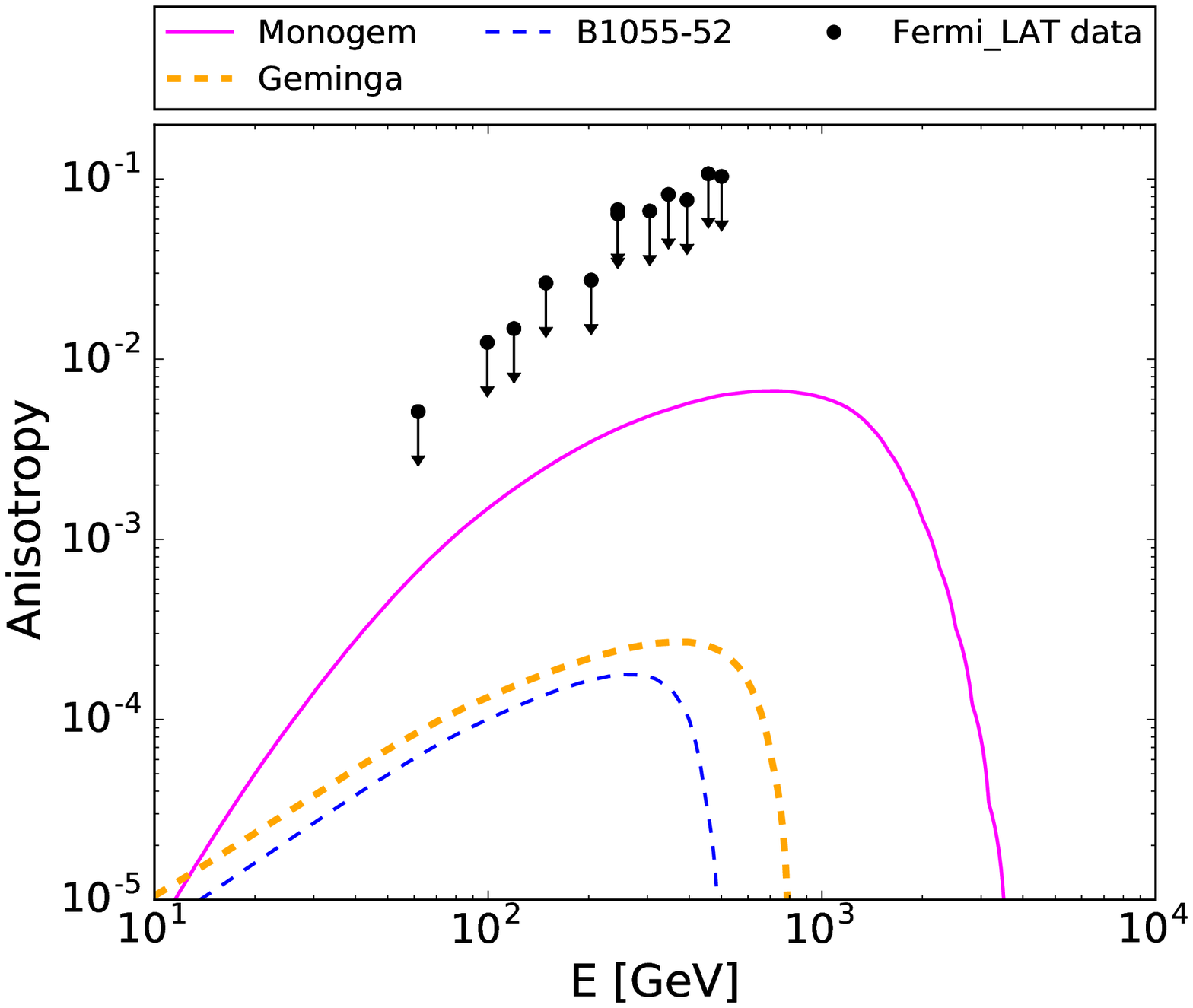}
\caption{{\em Left panel:} Positron fraction based on our model, for an injection time 85 kyr, (solid line) plotted against data from PAMELA~\cite{pamela_13_pos}, Fermi-LAT~\cite{fermi_pos_excess_12} and AMS-02~\cite{ams_02_positronfr}.
{\em Right panel:} Cosmic-ray electron and positron anisotropy for nearby pulsars in comparison with the Fermi-LAT upper limits~\cite{fermi_aniso}, for an injection time 85 kyr.}
\label{fig:anfrac}
\end{figure}

In our Fig. \ref{fig:leptons} and Fig. \ref{fig:inj_ref}, the contribution of PWN of nearby pulsars (as shown in Table \ref{tab:pulcat}) to the $e^{\pm}$ pair emission has been estimated for a diffusion coefficient
 $2.8 \times 10^{28} (E_e^{\pm}/4.5 \rm GeV)^{0.45} \rm cm^2/s$. This diffusion coefficient is exactly the same as the average diffusion coefficient used in the DRAGON code for the propagation of particles in the ISM.
We fix this diffusion coefficient and change the injection time $t_0$ and estimate the positron flux for these cases, which is shown in Figure \ref{fig:inj_ref} and in the right panel of Figure \ref{fig:leptons}. The contribution
of Monogem to the $e^{\pm}$ pair emission is effective for 50 kyr (left panel in Fig. \ref{fig:inj_ref}), and 85 kyr (right panel in Fig. \ref{fig:leptons}), while it is equal to zero for 200 kyr (right panel in Fig. \ref{fig:inj_ref}), due to its younger age.

In the left panel of Fig.~\ref{fig:anfrac} we have plotted the fractional positron flux of the total electron and positron flux. The solid (dashed) line corresponds to the flux model with (without) solar modulation taken into account.
Our model fits both the AMS-02 data~\cite{ams02_elec} and PAMELA data~\cite{pamela_13_pos} rather well.  Fermi-LAT data~\cite{fermi_pos_excess_12} in the figure show higher positron fraction  compared to PAMELA and AMS-02 data, in general.

\subsection{Anisotropy due to nearby discrete source}

The anisotropy of cosmic rays observed on Earth, which is mainly determined by the structure of the magnetic field in the solar neighbourhood, can be calculated using formalism in~\cite{berezinskii_buk} as $\frac{3D}{\upsilon} \frac{\nabla N}{N}$.  Here $\upsilon$ is the relativistic speed of the cosmic-ray particles, and $D$ is the diffusion coefficient for effective collision frequency $\sim \upsilon^2/D$ of cosmic-ray particles.  The anisotropy of cosmic-ray particles from a discrete source, below $\sim 100$ GeV can be washed out due to interplanetary magnetic field.  For nearby pulsars, required for our model, the main contribution to anisotropy occurs above $\sim 100$~GeV, which can be calculated as~\cite{berezinskii_buk,grasso_delta}
\begin{equation}
Anisotropy\, (E) = \frac{3r}{2 c (t-t_0)} 
\frac{(1-\delta)E/E_{\rm max}(t)}{1-[1-E/E_{\rm max}(t)]^{1-\delta}}
\frac{N_{e^-+e^+}^{\rm pulsar}(E)}{N_{e^-+e^+}^{\rm total}(E)}\,.       
\end{equation}
Here $t$ is the pulsar age, $r$ is the distance to the pulsar and $t_0$ is the injection time for pair emission from the pulsar.  The maximum energy $E_{\rm max}(t)=1/b_0(t-t_0)$, where $b_0=1.4 \times 10^{-16}$~GeV$^{-1}$~s$^{-1}$ is the energy loss rate for
electrons and positrons in the interstellar radiation field.  The $e^\pm$ pair emission ratio from a given nearby pulsar with respect to the total $e^\pm$ pair emission from all sources observe on
Earth, $N_{e^-+e^+}^{\rm pulsar}(E)/N_{e^-+e^+}^{\rm total}(E)$, decides the discrete source anisotropy. The recent AMS-02 data on $e^-+e^+$ and $e^+$ has been used to estimate the anisotropy for a nearby SNR and PWN \cite{donato_anisotropy}.

In the right panel of Fig.~\ref{fig:anfrac} we have plotted cosmic-ray $e^-+e^+$ anisotropy due to nearby pulsars.  As can be seen, the anisotropy in all cases is $\lesssim 1\%$ and much lower than the anisotropy upper limits measured by the Fermi-LAT~\cite{fermi_aniso}.

%%%%%%%%%%%%%%%%%%%%%%%%%%%%%%%%%%%%%%%%%%%%%%%%%%%%%
%%%%%%%%%%Section 4%%%%%%%%%%%%%
\section{Summary and Discussion}
%%%%%%%%%%%%%%%%%%%%%%%%%%%%%%%%
\label{sec:dissc}

We have modeled cosmic-ray nuclei and lepton flux data from AMS-02 using a Galactic population of conventional cosmi-ray nuclei and electron sources, a subdominant population of electron-positron sources and nearby pulsars producing electrons and positrons.  We have used the DRAGON code for propagation of cosmic rays in the Galaxy, as well as analytic diffusive loss equation for nearby pulsars.

The calculation of Boron to Carbon flux ratio using the DRAGON code constraints the Galactic halo size and the diffusion coefficient for the study of cosmic-ray diffusion in our Galaxy.  We fit both AMS-02 and PAMELA data very well.  In particular, we find a diffusion coefficient $2.8\times 10^{28}$ cm$^2$/s and an index 0.45.  This is closer to the Kraichnan turbulence spectral index and similar to what
was found in some of past studies (see, e.g.,~\cite{grasso_delta}) but different from the Kolmogorov turbulence spectrum with index 0.33. Such a difference in the index can arise due to different assumptions we have made in our
calculation about the Galaxy size, magnetic field structure, etc.  Also, we fit the B/C ratio data in the 1 GeV - 1 TeV range, while in \cite{bbyc_ams16} the ratio is fitted above 65 GV only. Furthermore, our choice of the diffusion index is also constrtained by requirements to fit the proton and Helium data simultaneously with the B/C ratio data. These results are listed in Table~\ref{tab:sumpar}.

The Proton flux we have calculated can fit the AMS-02 data very well but we could not reproduce the Helium data in the energy range below 4 GeV.  Our model also fits the PAMELA proton and Helium data above 4 GeV.  We believe that solar modulation effects is responsible for such discrepancies at low energies.
We found that the observed spectrum of electrons required two breaks for its successfull interpretation. The low energy break is due to electrons cooling in the Galactic magnetic field, while the second break, after which the flux is harder, may indicate a faster escape of high energy
electrons from their sources \cite{twobreaks}. In recent work, a break in the $e^-+e^+$ spectrum at 100 GeV has been seen, which should be caused by the electron spectrum due to its
dominating flux compared to positrons \cite{dmauro_100GeVbr}.

We needed an additional population of sources, distributed similar to the conventional source population but producing primary electron-positron fluxes.  A requirement for an additional source population to fit data is not new.  Indeed, plausible cosmic-ray sources in the spiral arms of the Galaxy been considered for the interpretation of positron flux and positron excess data using the DRAGON code~\cite{dragon_prd_14}.  The total power in $e^\pm$ fluxes from the additional source population in our case is $1.5\times 10^{53}$~GeV/Myr or only $5.4\%$ ($0.15\%$) of the power in primary electrons (protons) from the conventional sources.  Sources such as white dwarf pulsars~\cite{Usov:1993} with magnetic field $\gtrsim 10^9$~G (see, e.g.~\cite{Liebert:2002qu}) can accelerate $e^\pm$ to 10 TeV~\cite{Kashiyama:2010ui}.  Although the rotational powers of a white-dwarf pulsar and that of a neutron-star pulsar are similar, longevity of the former can be responsible for a much lower  total power injection in the Galaxy.  Other sources such as gamma-ray novae~\cite{Abdo:2010he} can also possibly accelerate particles to very-high energies~\cite{Razzaque:2010kp,Metzger:2015zka}.  In such a case the $e^\pm$ fluxes must come from $pp$ interactions, which may be responsible for the observed gamma rays from $\pi^0$ decays, and associated $\pi^\pm$ decays.  The primary proton flux in such as a case should be negligible compared to the conventional cosmic-ray sources.   

%In earlier studies of cosmic ray diffusion in our Galaxy, the importance of spiral arm cosmic ray sources has been considered for the interpretation of positron flux and excess using DRAGON code \cite{dragon_prd_14}. 

We have also used contributions to the $e^-$ and $e^+$ fluxes from the nearby pulsars, as is usually considered, e.g.,~\cite{venter_08,dragon_prd_14,hawc_hooper17} to explain observations in the $\gtrsim 100$~GeV-1~TeV range.  We have selected some of the sub-kpc bright pulsar candidates from the ATNF catalogue, shown in Table \ref{tab:pulcat}. Particles accelerated in pulsars are in general trapped in their PWN and injected into the interstellar medium when the pulsar wind merges into the ISM.  This process takes approximately 100 kyr after the pulsar birth period, e.g.,~\cite{venter_08}.
We found that the $e^\pm$ pair emission from nearby pulsars is very important for the interpretation of the positron fluxes in the energy range of 100-1000 GeV.
The flux from a pulsar can be optimized based on the injection time $t_0$ of $e^-$ and $e^+$ from its PWN, the diffusion coefficient for their propagation and its distance from us.

\begin{table}
\centering
%TABLE-I
\vskip 0.5cm
\caption{Parameters used in the study of cosmic ray diffusion in our Galaxy.}
\vskip 0.25cm
\begin{tabular}{|l|l|l|l|l|}
\hline
Physical parameters & Value \\
% & kpc& GeV & $GeV^{-1} cm^{-2} sec^{-1}$ \\
\hline
 Diffusion Coefficient $(D_0)$ & $2.8 \times 10^{28} ~\rm cm^2/\rm s$   \\
 Diffusion index ($\delta$) & 0.45  \\
 Low energy diffusion correction factor ($\eta$)&  -0.005  \\
 Diffusion coefficient scale height ($z_t$) &  4.0 kpc\\
 Alfven speed ($v_A$)& 19 km~$s^{-1}$  \\
 %Diffusion Coefficient, nearby source $(D_0)$ &$2.8 \times 10^{27} ~\rm cm^2/\rm s$  \\

\hline
\end{tabular}
\label{tab:sumpar}
\end{table}

Finally, we find that we can successfully model the positron-excess data using a sub-dominant cosmic-ray $e^\pm$ source population and a few nearby pulsars, along with positron flux from cosmic-ray nuclei interactions in ISM.
There is however some degeneracy in nearby pulsar contributions, mostly due
to an unknown time scale for injection of $e^\pm$ pairs.  Better knowledge of pulsar astrophysics and more precise data at high energies in future will be important to alleviate this degeneracy.

%The contribution of nearby pulsars to the positron fluxes can be seen explicitly in our Figure\ref{fig:inj_ref} and in the right panel of Figure \ref{fig:leptons}, and their electron contribution of same magnitude has also been included in our calculation. The parameters used in the making of these plots for injection time 10,000 years, 90,000 years and 200,000 are shown in Table \ref{tab:pul_feather}}.

\section*{Acknowledgements}

This research was supported by a GES fellowship to J.C.J. at the University of Johannesburg.  We thank C.~Evoli for providing us answers on our queries related to the DRAGON code, as well as A.~Harding and A.~Strong for discussions.
\label{sec:ack}

\bibliography{ref}

\providecommand{\href}[2]{#2}\begingroup\raggedright\begin{thebibliography}{10}

\bibitem{Barwick:1995gv}
S.~W. {Barwick}, J.~J. {Beatty}, C.~R. {Bower}, C.~{Chaput}, S.~{Coutu}, G.~{de
  Nolfo} et~al., \emph{{Cosmic Ray Positrons at High Energies: A New
  Measurement}},
  \href{http://dx.doi.org/10.1103/PhysRevLett.75.390}{\emph{Physical Review
  Letters} {\bf 75} (July, 1995) 390--393}.

\bibitem{Picozza:2006nm}
P.~{Picozza}, A.~M. {Galper}, G.~{Castellini}, O.~{Adriani}, F.~{Altamura},
  M.~{Ambriola} et~al., \emph{{PAMELA A payload for antimatter matter
  exploration and light-nuclei astrophysics}},
  \href{http://dx.doi.org/10.1016/j.astropartphys.2006.12.002}{\emph{Astroparticle
  Physics} {\bf 27} (Apr., 2007) 296--315},
  [\href{http://arxiv.org/abs/astro-ph/0608697}{{\tt astro-ph/0608697}}].

\bibitem{Atwood:2009ez}
W.~B. {Atwood}, A.~A. {Abdo}, M.~{Ackermann}, W.~{Althouse}, B.~{Anderson},
  M.~{Axelsson} et~al., \emph{{The Large Area Telescope on the Fermi Gamma-Ray
  Space Telescope Mission}},
  \href{http://dx.doi.org/10.1088/0004-637X/697/2/1071}{\emph{\apj} {\bf 697}
  (June, 2009) 1071--1102}, [\href{http://arxiv.org/abs/0902.1089}{{\tt
  0902.1089}}].

\bibitem{Aguilar:2013qda}
M.~{Aguilar}, G.~{Alberti}, B.~{Alpat}, A.~{Alvino}, G.~{Ambrosi}, K.~{Andeen}
  et~al., \emph{{First Result from the Alpha Magnetic Spectrometer on the
  International Space Station: Precision Measurement of the Positron Fraction
  in Primary Cosmic Rays of 0.5-350 GeV}},
  \href{http://dx.doi.org/10.1103/PhysRevLett.110.141102}{\emph{Physical Review
  Letters} {\bf 110} (Apr., 2013) 141102}.

\bibitem{Barwick:1997ig}
S.~W. {Barwick}, J.~J. {Beatty}, A.~{Bhattacharyya}, C.~R. {Bower}, C.~J.
  {Chaput}, S.~{Coutu} et~al., \emph{{Measurements of the Cosmic-Ray Positron
  Fraction from 1 to 50 GeV}},
  \href{http://dx.doi.org/10.1086/310706}{\emph{\apjl} {\bf 482} (June, 1997)
  L191--L194}, [\href{http://arxiv.org/abs/astro-ph/9703192}{{\tt
  astro-ph/9703192}}].

\bibitem{pamela_13_pos}
O.~{Adriani}, G.~C. {Barbarino}, G.~A. {Bazilevskaya}, R.~{Bellotti},
  A.~{Bianco}, M.~{Boezio} et~al., \emph{{Cosmic-Ray Positron Energy Spectrum
  Measured by PAMELA}},
  \href{http://dx.doi.org/10.1103/PhysRevLett.111.081102}{\emph{Physical Review
  Letters} {\bf 111} (Aug., 2013) 081102},
  [\href{http://arxiv.org/abs/1308.0133}{{\tt 1308.0133}}].

\bibitem{fermi_pos_excess_12}
M.~{Ackermann}, M.~{Ajello}, A.~{Allafort}, W.~B. {Atwood}, L.~{Baldini},
  G.~{Barbiellini} et~al., \emph{{Measurement of Separate Cosmic-Ray Electron
  and Positron Spectra with the Fermi Large Area Telescope}},
  \href{http://dx.doi.org/10.1103/PhysRevLett.108.011103}{\emph{Physical Review
  Letters} {\bf 108} (Jan., 2012) 011103},
  [\href{http://arxiv.org/abs/1109.0521}{{\tt 1109.0521}}].

\bibitem{Cholis:2008hb}
I.~{Cholis}, L.~{Goodenough}, D.~{Hooper}, M.~{Simet} and N.~{Weiner},
  \emph{{High energy positrons from annihilating dark matter}},
  \href{http://dx.doi.org/10.1103/PhysRevD.80.123511}{\emph{\prd} {\bf 80}
  (Dec., 2009) 123511}, [\href{http://arxiv.org/abs/0809.1683}{{\tt
  0809.1683}}].

\bibitem{Bergstrom:2008gr}
L.~{Bergstr{\"o}m}, T.~{Bringmann} and J.~{Edsj{\"o}}, \emph{{New positron
  spectral features from supersymmetric dark matter: A way to explain the
  PAMELA data?}},
  \href{http://dx.doi.org/10.1103/PhysRevD.78.103520}{\emph{\prd} {\bf 78}
  (Nov., 2008) 103520}, [\href{http://arxiv.org/abs/0808.3725}{{\tt
  0808.3725}}].

\bibitem{posexc_wdm}
T.~A. {Porter}, R.~P. {Johnson} and P.~W. {Graham}, \emph{{Dark Matter Searches
  with Astroparticle Data}},
  \href{http://dx.doi.org/10.1146/annurev-astro-081710-102528}{\emph{\araa}
  {\bf 49} (Sept., 2011) 155--194}, [\href{http://arxiv.org/abs/1104.2836}{{\tt
  1104.2836}}].

\bibitem{Cholis:2013psa}
I.~{Cholis} and D.~{Hooper}, \emph{{Dark matter and pulsar origins of the
  rising cosmic ray positron fraction in light of new data from the AMS}},
  \href{http://dx.doi.org/10.1103/PhysRevD.88.023013}{\emph{\prd} {\bf 88}
  (July, 2013) 023013}, [\href{http://arxiv.org/abs/1304.1840}{{\tt
  1304.1840}}].

\bibitem{hooper_ener_inj}
D.~{Hooper}, P.~{Blasi} and P.~{Dario Serpico}, \emph{{Pulsars as the sources
  of high energy cosmic ray positrons}},
  \href{http://dx.doi.org/10.1088/1475-7516/2009/01/025}{\emph{\jcap} {\bf 1}
  (Jan., 2009) 025}, [\href{http://arxiv.org/abs/0810.1527}{{\tt 0810.1527}}].

\bibitem{occamrazor_profumo}
S.~{Profumo}, \emph{{Dissecting cosmic-ray electron-positron data with Occam's
  razor: the role of known pulsars}},
  \href{http://dx.doi.org/10.2478/s11534-011-0099-z}{\emph{Central European
  Journal of Physics} {\bf 10} (Feb., 2012) 1--31},
  [\href{http://arxiv.org/abs/0812.4457}{{\tt 0812.4457}}].

\bibitem{Yuksel:2008rf}
H.~{Y{\"u}ksel}, M.~D. {Kistler} and T.~{Stanev}, \emph{{TeV Gamma Rays from
  Geminga and the Origin of the GeV Positron Excess}},
  \href{http://dx.doi.org/10.1103/PhysRevLett.103.051101}{\emph{Physical Review
  Letters} {\bf 103} (July, 2009) 051101},
  [\href{http://arxiv.org/abs/0810.2784}{{\tt 0810.2784}}].

\bibitem{Venter:2015gga}
C.~{Venter}, A.~{Kopp}, A.~K. {Harding}, P.~L. {Gonthier} and
  I.~{B{\"u}sching}, \emph{{Cosmic-ray Positrons from Millisecond Pulsars}},
  \href{http://dx.doi.org/10.1088/0004-637X/807/2/130}{\emph{\apj} {\bf 807}
  (July, 2015) 130}, [\href{http://arxiv.org/abs/1506.01211}{{\tt
  1506.01211}}].

\bibitem{pulsr_intr_zhang}
J.~{Feng} and H.-H. {Zhang}, \emph{{Pulsar interpretation of lepton spectra
  measured by AMS-02}},
  \href{http://dx.doi.org/10.1140/epjc/s10052-016-4092-y}{\emph{European
  Physical Journal C} {\bf 76} (May, 2016) 229},
  [\href{http://arxiv.org/abs/1504.03312}{{\tt 1504.03312}}].

\bibitem{posex_possintr}
Y.-Z. {Fan}, B.~{Zhang} and J.~{Chang}, \emph{{Electron/positron Excesses in
  the Cosmic Ray Spectrum and Possible Interpretations}},
  \href{http://dx.doi.org/10.1142/S0218271810018268}{\emph{International
  Journal of Modern Physics D} {\bf 19} (Nov., 2010) 2011--2058},
  [\href{http://arxiv.org/abs/1008.4646}{{\tt 1008.4646}}].

\bibitem{lin_yuanxj}
S.-J. {Lin}, Q.~{Yuan} and X.-J. {Bi}, \emph{{Quantitative study of the AMS-02
  electron/positron spectra: Implications for pulsars and dark matter
  properties}}, \href{http://dx.doi.org/10.1103/PhysRevD.91.063508}{\emph{\prd}
  {\bf 91} (Mar., 2015) 063508}, [\href{http://arxiv.org/abs/1409.6248}{{\tt
  1409.6248}}].

\bibitem{Shaviv:2009bu}
N.~J. {Shaviv}, E.~{Nakar} and T.~{Piran}, \emph{{Inhomogeneity in Cosmic Ray
  Sources as the Origin of the Electron Spectrum and the PAMELA Anomaly}},
  \href{http://dx.doi.org/10.1103/PhysRevLett.103.111302}{\emph{Physical Review
  Letters} {\bf 103} (Sept., 2009) 111302},
  [\href{http://arxiv.org/abs/0902.0376}{{\tt 0902.0376}}].

\bibitem{Kachelriess:2011qv}
M.~{Kachelrie{\ss}}, S.~{Ostapchenko} and R.~{Tom{\`a}s}, \emph{{Antimatter
  Production in Supernova Remnants}},
  \href{http://dx.doi.org/10.1088/0004-637X/733/2/119}{\emph{\apj} {\bf 733}
  (June, 2011) 119}, [\href{http://arxiv.org/abs/1103.5765}{{\tt 1103.5765}}].

\bibitem{hawc_hooper17}
D.~{Hooper}, I.~{Cholis}, T.~{Linden} and K.~{Fang}, \emph{{HAWC Observations
  Strongly Favor Pulsar Interpretations of the Cosmic-Ray Positron Excess}},
  {\emph{ArXiv e-prints} (Feb., 2017) },
  [\href{http://arxiv.org/abs/1702.08436}{{\tt 1702.08436}}].

\bibitem{blasi_posi_excess}
P.~{Blasi}, \emph{{Origin of the Positron Excess in Cosmic Rays}},
  \href{http://dx.doi.org/10.1103/PhysRevLett.103.051104}{\emph{Physical Review
  Letters} {\bf 103} (July, 2009) 051104},
  [\href{http://arxiv.org/abs/0903.2794}{{\tt 0903.2794}}].

\bibitem{subir_nsnr}
P.~{Mertsch} and S.~{Sarkar}, \emph{{AMS-02 data confront acceleration of
  cosmic ray secondaries in nearby sources}},
  \href{http://dx.doi.org/10.1103/PhysRevD.90.061301}{\emph{\prd} {\bf 90}
  (Sept., 2014) 061301}, [\href{http://arxiv.org/abs/1402.0855}{{\tt
  1402.0855}}].

\bibitem{Abeysekara:2017hyn}
A.~U. {Abeysekara}, A.~{Albert}, R.~{Alfaro}, C.~{Alvarez}, J.~D.
  {{\'A}lvarez}, R.~{Arceo} et~al., \emph{{The 2HWC HAWC Observatory Gamma Ray
  Catalog}}, {\emph{ArXiv e-prints} (Feb., 2017) },
  [\href{http://arxiv.org/abs/1702.02992}{{\tt 1702.02992}}].

\bibitem{gaggero2012cosmic}
D.~Gaggero, \emph{Cosmic Ray Diffusion in the Galaxy and Diffuse Gamma
  Emission}.
\newblock Springer Theses. Springer Berlin Heidelberg, 2012.

\bibitem{bbyc_ams16}
M.~{Aguilar}, \emph{Precision measurement of the boron to carbon flux ratio in
  cosmic rays from 1.9 gv to 2.6 tv with the alpha magnetic spectrometer on the
  international space station},
  \href{http://dx.doi.org/10.1103/PhysRevLett.117.231102}{\emph{Phys. Rev.
  Lett.} {\bf 117} (Nov, 2016) 231102}.

\bibitem{pamela_exper_one}
O.~{Adriani}, G.~C. {Barbarino}, G.~A. {Bazilevskaya}, R.~{Bellotti},
  M.~{Boezio}, E.~A. {Bogomolov} et~al., \emph{{New Measurement of the
  Antiproton-to-Proton Flux Ratio up to 100 GeV in the Cosmic Radiation}},
  \href{http://dx.doi.org/10.1103/PhysRevLett.102.051101}{\emph{Physical Review
  Letters} {\bf 102} (Feb., 2009) 051101},
  [\href{http://arxiv.org/abs/0810.4994}{{\tt 0810.4994}}].

\bibitem{pamela_exper_two}
O.~{Adriani}, G.~C. {Barbarino}, G.~A. {Bazilevskaya}, R.~{Bellotti},
  M.~{Boezio}, E.~A. {Bogomolov} et~al., \emph{{Cosmic-Ray Electron Flux
  Measured by the PAMELA Experiment between 1 and 625 GeV}},
  \href{http://dx.doi.org/10.1103/PhysRevLett.106.201101}{\emph{Physical Review
  Letters} {\bf 106} (May, 2011) 201101},
  [\href{http://arxiv.org/abs/1103.2880}{{\tt 1103.2880}}].

\bibitem{bbycpamela_14}
O.~{Adriani}, G.~C. {Barbarino}, G.~A. {Bazilevskaya}, R.~{Bellotti},
  M.~{Boezio}, E.~A. {Bogomolov} et~al., \emph{{Measurement of Boron and Carbon
  Fluxes in Cosmic Rays with the PAMELA Experiment}},
  \href{http://dx.doi.org/10.1088/0004-637X/791/2/93}{\emph{\apj} {\bf 791}
  (Aug., 2014) 93}, [\href{http://arxiv.org/abs/1407.1657}{{\tt 1407.1657}}].

\bibitem{Evoli:2016xgn}
C.~{Evoli}, D.~{Gaggero}, A.~{Vittino}, G.~{Di Bernardo}, M.~{Di Mauro},
  A.~{Ligorini} et~al., \emph{{Cosmic-ray propagation with DRAGON2: I.
  numerical solver and astrophysical ingredients}},
  \href{http://dx.doi.org/10.1088/1475-7516/2017/02/015}{\emph{\jcap} {\bf 2}
  (Feb., 2017) 015}, [\href{http://arxiv.org/abs/1607.07886}{{\tt
  1607.07886}}].

\bibitem{atoyan_ssc}
A.~M. {Atoyan}, F.~A. {Aharonian} and H.~J. {V{\"o}lk}, \emph{{Electrons and
  positrons in the galactic cosmic rays}},
  \href{http://dx.doi.org/10.1103/PhysRevD.52.3265}{\emph{\prd} {\bf 52}
  (Sept., 1995) 3265--3275}.

\bibitem{bell1_78}
A.~R. {Bell}, \emph{{The acceleration of cosmic rays in shock fronts. I}},
  \href{http://dx.doi.org/10.1093/mnras/182.2.147}{\emph{\mnras} {\bf 182}
  (Jan., 1978) 147--156}.

\bibitem{bell2_78}
A.~R. {Bell}, \emph{{The acceleration of cosmic rays in shock fronts. II}},
  \href{http://dx.doi.org/10.1093/mnras/182.3.443}{\emph{\mnras} {\bf 182}
  (Feb., 1978) 443--455}.

\bibitem{blandfr_chlr_87}
R.~{Blandford} and D.~{Eichler}, \emph{{Particle acceleration at astrophysical
  shocks: A theory of cosmic ray origin}},
  \href{http://dx.doi.org/10.1016/0370-1573(87)90134-7}{\emph{\physrep} {\bf
  154} (Oct., 1987) 1--75}.

\bibitem{rv_smp2007}
A.~W. {Strong}, I.~V. {Moskalenko} and V.~S. {Ptuskin}, \emph{{Cosmic-Ray
  Propagation and Interactions in the Galaxy}},
  \href{http://dx.doi.org/10.1146/annurev.nucl.57.090506.123011}{\emph{Annual
  Review of Nuclear and Particle Science} {\bf 57} (nov, 2007) 285--327},
  [\href{http://arxiv.org/abs/astro-ph/0701517}{{\tt astro-ph/0701517}}].

\bibitem{gas_model_galprop}
A.~W. {Strong}, I.~V. {Moskalenko}, O.~{Reimer}, S.~{Digel} and R.~{Diehl},
  \emph{{The distribution of cosmic-ray sources in the Galaxy, {$\gamma$}-rays
  and the gradient in the CO-to-H$_{2}$ relation}},
  \href{http://dx.doi.org/10.1051/0004-6361:20040172}{\emph{\aap} {\bf 422}
  (July, 2004) L47--L50}, [\href{http://arxiv.org/abs/astro-ph/0405275}{{\tt
  astro-ph/0405275}}].

\bibitem{pshirkov11}
M.~S. {Pshirkov}, P.~G. {Tinyakov}, P.~P. {Kronberg} and K.~J. {Newton-McGee},
  \emph{{Deriving the Global Structure of the Galactic Magnetic Field from
  Faraday Rotation Measures of Extragalactic Sources}},
  \href{http://dx.doi.org/10.1088/0004-637X/738/2/192}{\emph{\apj} {\bf 738}
  (Sept., 2011) 192}, [\href{http://arxiv.org/abs/1103.0814}{{\tt 1103.0814}}].

\bibitem{bernardo_aniso}
G.~{di Bernardo}, C.~{Evoli}, D.~{Gaggero}, D.~{Grasso}, L.~{Maccione} and
  M.~N. {Mazziotta}, \emph{{Implications of the cosmic ray electron spectrum
  and anisotropy measured with Fermi-LAT}},
  \href{http://dx.doi.org/10.1016/j.astropartphys.2010.11.005}{\emph{Astroparticle
  Physics} {\bf 34} (Feb., 2011) 528--538},
  [\href{http://arxiv.org/abs/1010.0174}{{\tt 1010.0174}}].

\bibitem{bbycratiosupp}
A.~W. {Strong} and I.~V. {Moskalenko}, \emph{{Propagation of Cosmic-Ray
  Nucleons in the Galaxy}}, \href{http://dx.doi.org/10.1086/306470}{\emph{\apj}
  {\bf 509} (Dec., 1998) 212--228},
  [\href{http://arxiv.org/abs/astro-ph/9807150}{{\tt astro-ph/9807150}}].

\bibitem{index_ref}
A.~{Castellina} and F.~{Donato}, \emph{{Diffusion coefficient and acceleration
  spectrum from direct measurements of charged cosmic ray nuclei}},
  \href{http://dx.doi.org/10.1016/j.astropartphys.2005.06.006}{\emph{Astroparticle
  Physics} {\bf 24} (Sept., 2005) 146--159},
  [\href{http://arxiv.org/abs/astro-ph/0504149}{{\tt astro-ph/0504149}}].

\bibitem{glee_axf_solmod}
L.~J. {Gleeson} and W.~I. {Axford}, \emph{{Solar Modulation of Galactic Cosmic
  Rays}}, \href{http://dx.doi.org/10.1086/149822}{\emph{\apj} {\bf 154} (Dec.,
  1968) 1011}.

\bibitem{lorimer06}
D.~R. {Lorimer}, A.~J. {Faulkner}, A.~G. {Lyne}, R.~N. {Manchester},
  M.~{Kramer}, M.~A. {McLaughlin} et~al., \emph{{The Parkes Multibeam Pulsar
  Survey - VI. Discovery and timing of 142 pulsars and a Galactic population
  analysis}},
  \href{http://dx.doi.org/10.1111/j.1365-2966.2006.10887.x}{\emph{\mnras} {\bf
  372} (Oct., 2006) 777--800},
  [\href{http://arxiv.org/abs/astro-ph/0607640}{{\tt astro-ph/0607640}}].

\bibitem{twobreaks}
A.~W. {Strong}, E.~{Orlando} and T.~R. {Jaffe}, \emph{{The interstellar
  cosmic-ray electron spectrum from synchrotron radiation and direct
  measurements}},
  \href{http://dx.doi.org/10.1051/0004-6361/201116828}{\emph{\aap} {\bf 534}
  (Oct., 2011) A54}, [\href{http://arxiv.org/abs/1108.4822}{{\tt 1108.4822}}].

\bibitem{database}
D.~{Maurin}, F.~{Melot} and R.~{Taillet}, \emph{{A database of charged cosmic
  rays}}, \href{http://dx.doi.org/10.1051/0004-6361/201321344}{\emph{\aap} {\bf
  569} (Sept., 2014) A32}, [\href{http://arxiv.org/abs/1302.5525}{{\tt
  1302.5525}}].

\bibitem{pamela_proton}
O.~{Adriani}, G.~C. {Barbarino}, G.~A. {Bazilevskaya}, R.~{Bellotti},
  M.~{Boezio}, E.~A. {Bogomolov} et~al., \emph{{PAMELA Measurements of
  Cosmic-Ray Proton and Helium Spectra}},
  \href{http://dx.doi.org/10.1126/science.1199172}{\emph{Science} {\bf 332}
  (Apr., 2011) 69}, [\href{http://arxiv.org/abs/1103.4055}{{\tt 1103.4055}}].

\bibitem{ams_proton}
M.~{Aguilar}, D.~{Aisa}, B.~{Alpat}, A.~{Alvino}, G.~{Ambrosi}, K.~{Andeen}
  et~al., \emph{{Precision Measurement of the Proton Flux in Primary Cosmic
  Rays from Rigidity 1 GV to 1.8 TV with the Alpha Magnetic Spectrometer on the
  International Space Station}},
  \href{http://dx.doi.org/10.1103/PhysRevLett.114.171103}{\emph{Physical Review
  Letters} {\bf 114} (May, 2015) 171103}.

\bibitem{heams2}
M.~{Aguilar}, D.~{Aisa}, B.~{Alpat}, A.~{Alvino}, G.~{Ambrosi}, K.~{Andeen}
  et~al., \emph{{Precision Measurement of the Helium Flux in Primary Cosmic
  Rays of Rigidities 1.9 GV to 3 TV with the Alpha Magnetic Spectrometer on the
  International Space Station}},
  \href{http://dx.doi.org/10.1103/PhysRevLett.115.211101}{\emph{Physical Review
  Letters} {\bf 115} (Nov., 2015) 211101}.

\bibitem{ruderman_86_pcap}
K.~S. {Cheng}, C.~{Ho} and M.~{Ruderman}, \emph{{Energetic radiation from
  rapidly spinning pulsars. I - Outer magnetosphere gaps. II - VELA and Crab}},
  \href{http://dx.doi.org/10.1086/163829}{\emph{\apj} {\bf 300} (Jan., 1986)
  500--539}.

\bibitem{baring_pcpa}
M.~G. {Baring}, \emph{{High-energy emission from pulsars: the polar cap
  scenario}}, \href{http://dx.doi.org/10.1016/j.asr.2003.08.020}{\emph{Advances
  in Space Research} {\bf 33} (2004) 552--560},
  [\href{http://arxiv.org/abs/astro-ph/0308296}{{\tt astro-ph/0308296}}].

\bibitem{blasi_40_50kyr}
P.~{Blasi} and E.~{Amato}, \emph{{Positrons from pulsar winds}},
  {\emph{Astrophysics and Space Science Proceedings} {\bf 21} (2011) 624},
  [\href{http://arxiv.org/abs/1007.4745}{{\tt 1007.4745}}].

\bibitem{venter_08}
I.~{B{\"u}sching}, O.~C. {de Jager}, M.~S. {Potgieter} and C.~{Venter},
  \emph{{A Cosmic-Ray Positron Anisotropy due to Two Middle-Aged, Nearby
  Pulsars?}}, \href{http://dx.doi.org/10.1086/588465}{\emph{\apjl} {\bf 678}
  (May, 2008) L39}, [\href{http://arxiv.org/abs/0804.0220}{{\tt 0804.0220}}].

\bibitem{displs_min_1}
J.~P.~W. {Verbiest}, J.~M. {Weisberg}, A.~A. {Chael}, K.~J. {Lee} and D.~R.
  {Lorimer}, \emph{{On Pulsar Distance Measurements and Their Uncertainties}},
  \href{http://dx.doi.org/10.1088/0004-637X/755/1/39}{\emph{\apj} {\bf 755}
  (Aug., 2012) 39}, [\href{http://arxiv.org/abs/1206.0428}{{\tt 1206.0428}}].

\bibitem{b105552dis_2}
R.~P. {Mignani}, G.~G. {Pavlov} and O.~{Kargaltsev}, \emph{{Optical-Ultraviolet
  Spectrum and Proper Motion of the Middle-aged Pulsar B1055-52}},
  \href{http://dx.doi.org/10.1088/0004-637X/720/2/1635}{\emph{\apj} {\bf 720}
  (Sept., 2010) 1635--1643}, [\href{http://arxiv.org/abs/1007.2940}{{\tt
  1007.2940}}].

\bibitem{taylr_cordes_93}
J.~H. {Taylor} and J.~M. {Cordes}, \emph{{Pulsar distances and the galactic
  distribution of free electrons}},
  \href{http://dx.doi.org/10.1086/172870}{\emph{\apj} {\bf 411} (July, 1993)
  674--684}.

\bibitem{cordes_lazio}
J.~M. {Cordes} and T.~J.~W. {Lazio}, \emph{{NE2001.I. A New Model for the
  Galactic Distribution of Free Electrons and its Fluctuations}}, {\emph{ArXiv
  Astrophysics e-prints} (July, 2002) },
  [\href{http://arxiv.org/abs/astro-ph/0207156}{{\tt astro-ph/0207156}}].

\bibitem{lingenfelter_69}
R.~E. {Lingenfelter}, \emph{{Pulsars and Local Cosmic Ray Prehistory}},
  \href{http://dx.doi.org/10.1038/2241182a0}{\emph{\nat} {\bf 224} (Dec., 1969)
  1182--1186}.

\bibitem{blu_gould1970}
G.~R. {Blumenthal} and R.~J. {Gould}, \emph{{Bremsstrahlung, Synchrotron
  Radiation, and Compton Scattering of High-Energy Electrons Traversing Dilute
  Gases}}, \href{http://dx.doi.org/10.1103/RevModPhys.42.237}{\emph{Reviews of
  Modern Physics} {\bf 42} (1970) 237--271}.

\bibitem{kobayashi_10gev}
T.~{Kobayashi}, Y.~{Komori}, K.~{Yoshida} and J.~{Nishimura}, \emph{{The Most
  Likely Sources of High-Energy Cosmic-Ray Electrons in Supernova Remnants}},
  \href{http://dx.doi.org/10.1086/380431}{\emph{\apj} {\bf 601} (Jan., 2004)
  340--351}, [\href{http://arxiv.org/abs/astro-ph/0308470}{{\tt
  astro-ph/0308470}}].

\bibitem{berezinskii_buk}
V.~S. {Berezinskii}, S.~V. {Bulanov}, V.~A. {Dogiel} and V.~S. {Ptuskin},
  \emph{{Astrophysics of cosmic rays}}.
\newblock 1990.

\bibitem{shen_pulsar70}
C.~S. {Shen}, \emph{{Pulsars and Very High-Energy Cosmic-Ray Electrons}},
  \href{http://dx.doi.org/10.1086/180650}{\emph{\apjl} {\bf 162} (Dec., 1970)
  L181}.

\bibitem{malyshev_pulsarageetc}
D.~{Malyshev}, I.~{Cholis} and J.~{Gelfand}, \emph{{Pulsars versus dark matter
  interpretation of ATIC/PAMELA}},
  \href{http://dx.doi.org/10.1103/PhysRevD.80.063005}{\emph{\prd} {\bf 80}
  (Sept., 2009) 063005}, [\href{http://arxiv.org/abs/0903.1310}{{\tt
  0903.1310}}].

\bibitem{grasso_delta}
D.~{Grasso}, S.~{Profumo}, A.~W. {Strong}, L.~{Baldini}, R.~{Bellazzini}, E.~D.
  {Bloom} et~al., \emph{{On possible interpretations of the high energy
  electron-positron spectrum measured by the Fermi Large Area Telescope}},
  \href{http://dx.doi.org/10.1016/j.astropartphys.2009.07.003}{\emph{Astroparticle
  Physics} {\bf 32} (Sept., 2009) 140--151},
  [\href{http://arxiv.org/abs/0905.0636}{{\tt 0905.0636}}].

\bibitem{ams02_elec}
M.~{Aguilar}, D.~{Aisa}, A.~{Alvino}, G.~{Ambrosi}, K.~{Andeen}, L.~{Arruda}
  et~al., \emph{{Electron and Positron Fluxes in Primary Cosmic Rays Measured
  with the Alpha Magnetic Spectrometer on the International Space Station}},
  \href{http://dx.doi.org/10.1103/PhysRevLett.113.121102}{\emph{Physical Review
  Letters} {\bf 113} (Sept., 2014) 121102}.

\bibitem{ams_02_positronfr}
L.~{Accardo}, M.~{Aguilar}, D.~{Aisa}, A.~{Alvino}, G.~{Ambrosi}, K.~{Andeen}
  et~al., \emph{{High Statistics Measurement of the Positron Fraction in
  Primary Cosmic Rays of 0.5-500 GeV with the Alpha Magnetic Spectrometer on
  the International Space Station}},
  \href{http://dx.doi.org/10.1103/PhysRevLett.113.121101}{\emph{Physical Review
  Letters} {\bf 113} (Sept., 2014) 121101}.

\bibitem{fermi_aniso}
M.~{Ackermann}, M.~{Ajello}, W.~B. {Atwood}, L.~{Baldini}, J.~{Ballet},
  G.~{Barbiellini} et~al., \emph{{Searches for cosmic-ray electron anisotropies
  with the Fermi Large Area Telescope}},
  \href{http://dx.doi.org/10.1103/PhysRevD.82.092003}{\emph{\prd} {\bf 82}
  (Nov., 2010) 092003}, [\href{http://arxiv.org/abs/1008.5119}{{\tt
  1008.5119}}].

\bibitem{donato_anisotropy}
S.~{Manconi}, M.~{Di Mauro} and F.~{Donato}, \emph{{Dipole anisotropy in cosmic
  electrons and positrons: inspection on local sources}},
  \href{http://dx.doi.org/10.1088/1475-7516/2017/01/006}{\emph{\jcap} {\bf 1}
  (Jan., 2017) 006}, [\href{http://arxiv.org/abs/1611.06237}{{\tt
  1611.06237}}].

\bibitem{dmauro_100GeVbr}
M.~{Di Mauro}, S.~{Manconi}, A.~{Vittino}, F.~{Donato}, N.~{Fornengo},
  L.~{Baldini} et~al., \emph{{Theoretical interpretation of Pass 8 $Fermi$-LAT
  $e^+ + e^-$ data}}, {\emph{ArXiv e-prints} (Mar., 2017) },
  [\href{http://arxiv.org/abs/1703.00460}{{\tt 1703.00460}}].

\bibitem{dragon_prd_14}
D.~{Gaggero}, L.~{Maccione}, D.~{Grasso}, G.~{Di Bernardo} and C.~{Evoli},
  \emph{{PAMELA and AMS-02 e$^{+}$ and e$^{-}$ spectra are reproduced by
  three-dimensional cosmic-ray modeling}},
  \href{http://dx.doi.org/10.1103/PhysRevD.89.083007}{\emph{\prd} {\bf 89}
  (Apr., 2014) 083007}, [\href{http://arxiv.org/abs/1311.5575}{{\tt
  1311.5575}}].

\bibitem{Usov:1993}
V.~V. {Usov}, \emph{{High-frequency emission of X-ray pulsar 1E 2259+586}},
  \href{http://dx.doi.org/10.1086/172792}{\emph{\apj} {\bf 410} (June, 1993)
  761--763}.

\bibitem{Liebert:2002qu}
J.~{Liebert}, P.~{Bergeron} and J.~B. {Holberg}, \emph{{The True Incidence of
  Magnetism Among Field White Dwarfs}},
  \href{http://dx.doi.org/10.1086/345573}{\emph{\aj} {\bf 125} (Jan., 2003)
  348--353}, [\href{http://arxiv.org/abs/astro-ph/0210319}{{\tt
  astro-ph/0210319}}].

\bibitem{Kashiyama:2010ui}
K.~{Kashiyama}, K.~{Ioka} and N.~{Kawanaka}, \emph{{White dwarf pulsars as
  possible cosmic ray electron-positron factories}},
  \href{http://dx.doi.org/10.1103/PhysRevD.83.023002}{\emph{\prd} {\bf 83}
  (Jan., 2011) 023002}, [\href{http://arxiv.org/abs/1009.1141}{{\tt
  1009.1141}}].

\bibitem{Abdo:2010he}
A.~A. {Abdo}, M.~{Ackermann}, M.~{Ajello}, W.~B. {Atwood}, L.~{Baldini},
  J.~{Ballet} et~al., \emph{{Gamma-Ray Emission Concurrent with the Nova in the
  Symbiotic Binary V407 Cygni}},
  \href{http://dx.doi.org/10.1126/science.1192537}{\emph{Science} {\bf 329}
  (Aug., 2010) 817--821}, [\href{http://arxiv.org/abs/1008.3912}{{\tt
  1008.3912}}].

\bibitem{Razzaque:2010kp}
S.~{Razzaque}, P.~{Jean} and O.~{Mena}, \emph{{High energy neutrinos from novae
  in symbiotic binaries: The case of V407 Cygni}},
  \href{http://dx.doi.org/10.1103/PhysRevD.82.123012}{\emph{\prd} {\bf 82}
  (Dec., 2010) 123012}, [\href{http://arxiv.org/abs/1008.5193}{{\tt
  1008.5193}}].

\bibitem{Metzger:2015zka}
B.~D. {Metzger}, D.~{Caprioli}, I.~{Vurm}, A.~M. {Beloborodov}, I.~{Bartos} and
  A.~{Vlasov}, \emph{{Novae as Tevatrons: prospects for CTA and IceCube}},
  \href{http://dx.doi.org/10.1093/mnras/stw123}{\emph{\mnras} {\bf 457} (Apr.,
  2016) 1786--1795}, [\href{http://arxiv.org/abs/1510.07639}{{\tt
  1510.07639}}].

\end{thebibliography}\endgroup
\bibliographystyle{JHEP}

%\label{lastpage}
\end{document}